\documentclass[12pt,letter]{article}

\usepackage{graphicx, epsfig, color}
\def\eslt{\not\!\!{E_T}}
\def\to{\rightarrow}

\def\bi{\begin{itemize}}
\def\ei{\end{itemize}}
\def\te{\tilde e}

\def\tu{\tilde u}
\def\sps1ap{SPS1a$^\prime$}
\def\c1p{C1$^\prime$}

\def\tb{\tilde b}

\def\tst{\tilde t}
\def\ttau{\tilde \tau}

\def\tg{\tilde g}
\def\tnu{\tilde\nu}

\def\tq{\tilde q}

\def\tw{\widetilde W}
\def\tz{\widetilde Z}
\def\alt{\stackrel{<}{\sim}}
\def\agt{\stackrel{>}{\sim}}
\def\be{\begin{equation}}  
\def\ee{\end{equation}}  
\def\bea{\begin{eqnarray}}  
\def\eea{\end{eqnarray}}  
\def\beas{\begin{eqnarray*}}  
\def\eeas{\end{eqnarray*}}  
\newcommand\prd[3]{{\it Phys.\ Rev.\ }{\bf D #1} (#2) #3}

\newcommand\plb[3]{{\it Phys.\ Lett.\ }{\bf B #1} (#2) #3}

\newcommand{\lfig}{10 cm}
\newcommand{\sfig}{8 cm}

\begin{document}
\begin{titlepage}
\begin{flushright}
UH-511-1269-16
\end{flushright}

\vspace{0.5cm}
\begin{center}
{\Large \bf Gluino reach and mass extraction at the LHC\\ 
in radiatively-driven natural SUSY
}\\ 
\vspace{1.2cm} \renewcommand{\thefootnote}{\fnsymbol{footnote}}
{\large Howard Baer$^1$\footnote[1]{Email: baer@nhn.ou.edu }, 
Vernon Barger$^2$\footnote[2]{Email: barger@pheno.wisc.edu },
James S. Gainer$^3$\footnote[3]{Email: jgainer@hawaii.edu },
Peisi Huang$^{4,5,6}$\footnote[4]{Email: huang@physics.tamu.edu },\\
Michael Savoy$^1$\footnote[5]{Email: savoy@nhn.ou.edu },
Dibyashree Sengupta$^1$\footnote[6]{Email: Dibyashree.Sengupta-1@ou.edu },
and Xerxes Tata$^3$\footnote[7]{Email: tata@phys.hawaii.edu }
}\\ 
\vspace{1.2cm} \renewcommand{\thefootnote}{\arabic{footnote}}
{\it 
$^1$Dept. of Physics and Astronomy,
University of Oklahoma, Norman, OK 73019, USA \\
}
{\it 
$^2$Dept. of Physics,
University of Wisconsin, Madison, WI 53706, USA \\
}
{\it 
$^3$Dept. of Physics,
University of Hawaii, Honolulu, HI 96822, USA \\
}
{\it 
$^4$Enrico Fermi Institute, University of Chicago,
Chicago, IL 60637, USA\\
}
{\it
$^5$HEP Division, Argonne National Laboratory,
9700 Cass Ave., Argonne, IL 60439, USA}
{\it
$^6$Mitchell Institute for Fundamental Physics and Astronomy,
Texas A\&M University, College Station, TX 77843, USA\\
}
\end{center}

\vspace{0.5cm}
\begin{abstract}
Radiatively-driven natural SUSY (RNS) models enjoy electroweak 
naturalness at the 10\% level while respecting LHC sparticle and Higgs
mass constraints. Gluino and top squark masses can range up to several
TeV (with other squarks even heavier) but a set of light Higgsinos are
required with mass not too far above $m_h\sim 125$ GeV.  Within the RNS
framework, gluinos dominantly decay via 
$\tg \to t\tst_1^{*},\ \bar{t}\tst_1 \to t\bar{t}\tz_{1,2}$ or 
$t\bar{b}\tw_1^-+c.c.$, 
where the decay products of the higgsino-like
$\tw_1$ and $\tz_2$ are very soft. Gluino pair production is, therefore,
signalled by events with up to four hard $b$-jets and large $\eslt$. We
devise a set of cuts to isolate a relatively pure gluino sample at the
(high luminosity) LHC and show that in the RNS model with  very heavy
squarks, the gluino signal will be accessible for $m_{\tg} < 2400 \ 
(2800)$~GeV for an integrated luminosity of 300 (3000)~fb$^{-1}$.  We
also show that the measurement of the rate of gluino events in the clean
sample mentioned above allows for a determination of $m_{\tg}$ with a
statistical precision of 2.5-5\% (depending on the integrated luminosity
and the gluino mass) over the range of gluino masses where a 5$\sigma$
discovery is possible at the LHC.
\noindent 
\vspace*{0.8cm}

\end{abstract}

\end{titlepage}

\section{Introduction}
\label{sec:intro}

Supersymmetric (SUSY) models of particle physics are strongly
motivated because they provide a solution to the gauge hierarchy
problem (GHP)~\cite{ghp} which arises when the spin zero Higgs sector of
the Standard Model (SM) is coupled to a high mass sector as, for
instance, in a Grand Unified Theory (GUT).  SUSY models are
indirectly supported by data in that (1) the measured values of the
three gauge coupling strengths unify in the minimal supersymmetrized
Standard Model (MSSM)~\cite{gauge}, (2) the top-quark mass is measured
to lie in the range required by SUSY to trigger a radiative breakdown
of electroweak symmetry~\cite{rewsb}, and (3) the measured value of the
Higgs boson mass~\cite{lhc_h} lies squarely within the range required by the
MSSM~\cite{mhiggs}.  However, so far no unambiguous signal for
superparticles has emerged from LHC searches~\cite{lhc_s}.  In the case
of the gluino, $\tg$, (the spin-$1/2$ superpartner of the gluon), recent
search results in the context of simplified models place mass limits
as high as $m_{\tg}\agt 1500-1900$ GeV~\cite{atlas_4b,ichep16} for a massless
lightest SUSY particle (LSP), depending on the assumed decay of the
gluino.  This may be contrasted with early expectations from
naturalness, such as the Barbieri-Giudice (BG) $\Delta_{\rm BG}<30$
bounds~\cite{bg}, which required $m_{\tg}\alt 350$ GeV.\footnote{
The BG results were originally presented for $\Delta_{BG}<10$ but we 
re-scale the results to $\Delta_{BG}<30$ for direct comparison with our 
upper limits for $\Delta_{EW}<30$. The onset of fine-tuning for 
$\Delta_{EW}\agt 30$ is visually displayed in Fig. 1 of Ref. \cite{upper}.}

%
%
Likewise,  LHC simplified model analyses typically
require $m_{\tst_1}\agt 700-850$ GeV~\cite{ichep16,lhc_stop} to  be
contrasted with Dimopoulos-Giudice (DG) $\Delta_{BG}<30$ 
fine-tuning bounds~\cite{dg} that
require $m_{\tst_1}\alt 500$ GeV or with Higgs mass large log
bounds~\cite{oldnsusy} which require {\it three} third generation
squarks of mass $\alt 500$ GeV.  Conflicts such as these 
have led many to question whether weak scale SUSY is indeed nature's
chosen solution to the GHP, or whether nature follows some entirely
different direction~\cite{crisis}.

A more conservative approach to naturalness has been adopted in Refs.
\cite{ltr,rns}. Here, one requires that there are no large cancellations 
among the various terms on the right hand side of the well-known expression
that yields the measured value of $m_Z$ in terms of the
weak scale SUSY Lagrangian parameters via the scalar potential
minimization condition:
\be
\frac{m_Z^2}{2}=\frac{m_{H_d}^2+\Sigma_d^d-(m_{H_u}^2+\Sigma_u^u)\tan^2\beta}{\tan^2\beta -1}-\mu^2 \simeq -m_{H_u}^2-\Sigma_u^u -\mu^2 .
\label{eq:mzs}
\ee
The $\Sigma_u^u$ and $\Sigma_d^d$ terms in Eq.~(\ref{eq:mzs}) arise
from 1-loop corrections to the scalar potential (expressions can be
found in the Appendix of Ref.~\cite{rns}), $m_{H_u}^2$ and $m_{H_d}^2$
are the soft SUSY breaking Higgs mass parameters, $\tan \beta \equiv
\langle H_u \rangle / \langle H_d \rangle$ is the ratio of the Higgs
field VEVs and $\mu$ is the superpotential (SUSY conserving)
Higgs/higgsino mass parameter.  SUSY models requiring large
cancellations between the various terms on the right-hand-side of
Eq.~(\ref{eq:mzs}) to reproduce the measured value of $m_Z^2$ are
regarded as unnatural, or fine-tuned.
%
Thus, the {\it electroweak} naturalness
measure, $\Delta_{\rm EW}$, is given by the maximum value of the ratio of each
term on the right-hand-side of\ Eq.~(\ref{eq:mzs}) and $m_Z^2/2$.  

This conservative approach to naturalness allows for the possibility
that high scale parameters that have been taken to be independent
will, in fact, turn out to be correlated once SUSY breaking is
understood. Ignoring these correlations will lead to an overestimate
of the UV sensitivity of a theory, making it appear to be fine-tuned.
In Ref.~\cite{comp3,mt,seige} it is argued that if all correlations among
parameters are correctly implemented the conventional
Barbieri-Giudice measure reduces to $\Delta_{\rm EW}$ and that a
high-scale theory that predicts these parameter correlations would be
natural. We urge using the more conservative electroweak measure for
discussions of naturalness since disregarding the possibility of
parameter correlations may lead to prematurely discarding what may be
a perfectly viable effective theory.

For SUSY models with electroweak naturalness, we have:
\begin{itemize}
\item $|\mu |\sim 100-300$~GeV (the closer to $m_Z$ the more natural)
  leading to the requirement of four light higgsinos $\tz_{1,2}$ and
  $\tw_1^\pm$ of similar mass values $\sim |\mu |$,

\item $m_{H_u}^2$ must be driven radiatively to small negative values
  $\sim -(100-300)^2$ GeV$^2$ at the weak scale (for this reason, these
  models are said to exhibit radiatively-driven naturalness, and have
  been dubbed radiatively-driven natural SUSY (RNS)~\cite{ltr,rns}).

\item the radiative corrections $\left|\Sigma_u^u(i)\right|\alt (100-300)^2$
  GeV$^2$. The largest of these typically arise from the top squarks
  and require for $\Delta_{\rm EW}<30$ that $m_{\tst_1}\alt 3$ TeV, a
  factor 10 higher than the aforementioned BG/DG bounds. Gluinos
  contribute to $\Sigma_u^u$ at two-loop order~\cite{sundrum} and in
  models with gaugino mass unification, then $m_{\tg}\alt 4$
  TeV~\cite{rns,upper,guts}.
\end{itemize}
We see that SUSY models with electroweak naturalness can easily
respect LHC sparticle mass bounds and are in accord with the measured value
of $m_h$, which requires large mixing among top squarks~\cite{h125}.
%

What of LHC signatures in RNS models? Old favorites like searches
for gluino pairs ($pp\to \tg\tg X$ where $X$ represents assorted
hadronic debris) are still viable where now $\tg\to \tst_1 t$ if
kinematically allowed or $\tg\to t\bar{t}\tz_i$ or $t\bar{b}\tw_j$
when $m_{\tg}<m_{\tst_1}+m_t$~\cite{woodside}. In the case of RNS, the
$m_{\tg}-m_{\tz_1}$ mass gap is expected to be {\it larger} than in
models such as CMSSM/mSUGRA where the $\tz_1$ is typically bino-like.
In addition, for RNS, the $\tz_2$ produced in gluino cascade
decays leads to the presence of opposite-sign/same flavor dilepton
pairs with $m(\ell^+\ell^- )<m_{\tz_2}-m_{\tz_1}\sim 10-20$
GeV~\cite{lhc,baris}. However, new signatures also arise for RNS. Wino pair
production $pp\to \tw_2\tz_4$ can occur at large rates leading to the
low background same-sign diboson signature from $\tw_2\to
W^+\tz_{1,2}$ and $\tz_4\to W^+\tw_1^-$ decays~\cite{lhc, lhcltr}.  This
very clean signature leads to the greatest reach for SUSY in the
$m_{1/2}$ direction for an integrated luminosity $L\agt 100-200$
fb$^{-1}$. Also, direct higgsino pair production $pp\to\tz_1\tz_2 j$
followed by $\tz_2\to\ell^+\ell^-\tz_1$ decay offers substantial reach
in the $\mu$ direction of parameter space~\cite{z1z2j}.  Combined,
these latter two signatures offer high luminosity (HL) LHC a complete
coverage of RNS SUSY with unified gaugino masses 
for $\Delta_{\rm EW}<30$ and $L\sim 3000$ fb$^{-1}$~\cite{lhc2}. 
In addition to LHC searches, an International
Linear $e^+e^-$ Collider (ILC) with $\sqrt{s}\agt 500-600\ {\rm
  GeV}>2m(higgsino)$ would be a {\it higgsino factory} and completely
cover the $\Delta_{\rm EW} \le 30$ RNS parameter space and allow
 precision measurements that would serve to elucidate the natural
origin of $W$, $Z$ and Higgs boson masses~\cite{ilc,Baer:2016new}.

Although the discovery of the gluino at the LHC is not guaranteed over
the viable RNS parameter space, we re-examine gluino pair production
signatures expected within the RNS framework.  Our purpose is first, to
delineate the gluino reach of LHC14 and its high-luminosity upgrade, and
second, to study the extent to which the gluino mass may be extracted at
the LHC.  Although not required by naturalness, one usually takes 
first and second generation matter scalar mass parameters, assumed unified to a
value $m_0$ at scale $Q=m_{GUT}$, to be in the multi-TeV range. This
alleviates the SUSY flavour problem with little impact on naturalness as
long as these scalars satisfy well-motivated intra-generational
degeneracy patterns~\cite{maren}.
%
For integrated luminosities in excess of 100~fb$^{-1}$ that should be
accumulated within the next few years, we show that judicious cuts can
be found so that the gluino pair production signal emerges with very
little SM background in the data sample, allowing for a gluino reach
well beyond the expectation within the mSUGRA/CMSSM framework. Moreover,
assuming decoupled first and second generation squarks, the measured
event rate from the gluino signal depends {\it only} on the value of
$m_{\tg}$.  The rate for gluino events after cuts that 
eliminate most of the SM background can, therefore,
be used to extract the gluino mass, assuming that
gluino events as well as the experimental detector can be reliably
modeled. This ``counting rate'' method of extracting
$m_{\tg}$~\cite{gabe} has several advantages over the kinematic
methods which have been advocated~\cite{methods}.  It remains viable even
if a variety of complicated cascade decay topologies are expected to be
present. In addition, it is unaffected by ambiguities over which jets or
leptons are to be associated with which of the two gluinos that are
produced. We explore the counting rate extraction of $m_{\tg}$ in RNS
SUSY and find it typically leads to extraction of $m_{\tg}$ with a
statistical precision of 2.5-4\%, depending on the value of $m_{\tg}$ and
the assumed integrated luminosity.

The rest of this paper is organized as follows. 
In Sec.~\ref{sec:model} we present the RNS model line that we adopt 
for our analysis  and briefly describe the event topologies expected 
from gluino pair production within the RNS framework 
using a benchmark point with $m_{\tg}= 2$~TeV for illustration. 
In Sec.~\ref{sec:evgen}, we
discuss details of our simulation of the SUSY signal as well as the
relevant SM backgrounds. In Sec.~\ref{sec:cuts} we describe the
analysis cuts to select out gluino events from SM backgrounds and show
that it is possible to reduce the background level to no more 3\% for
our benchmark case. In Sec.~\ref{subsec:gluino_reach} we show our
projections for the mass reach for gluinos in the RNS framework, while
in Sec.~\ref{subsec:gluino_mass} we show the precision with which
$m_{\tg}$ may be extracted at the LHC. Finally, we summarize our
results in Sec.~\ref{sec:conc}.

\section{An RNS model line}
\label{sec:model}

To facilitate the examination of gluino signals in models with
natural SUSY spectra,
we adopt the RNS model-line first introduced in Ref.~\cite{lhc} (except
that we now take $\tan\beta=10$). Specifically, we work 
within the framework of the two extra parameter
non-universal Higgs model (NUHM2)~\cite{nuhm2} with parameter inputs,
\be
m_0,\ m_{1/2},\ A_0,\ \tan\beta ,\ \mu,\ m_A\ \ \ (NUHM2)\;.
\ee 
We use Isajet/Isasugra 7.85 spectrum generator~\cite{isajet} to obtain
sparticle masses.  For our model line, we adopt parameter choices
$m_0=5000$ GeV, $A_0=-8000$ GeV, $\tan\beta =10$, $\mu=150$ GeV and
$m_A=1000$ GeV, while $m_{1/2}$ varies across the range $600-1200$ GeV
corresponding to a gluino mass range of $m_{\tg}\sim 1600-2800$ GeV,
{\it i.e.}, starting just below present LHC bounds on $m_{\tg}$ and
extending just beyond the projected reach for HL-LHC.  The spectrum,
together with some low energy observables, is illustrated for a benchmark
point with $m_{\tg}\simeq 2000$ GeV in Table~\ref{tab:bm}.  Along this
model line, the computed value of the light Higgs mass is quite stable
and varies over $m_h:124.1-124.7$ GeV.  (We expect a couple GeV theory
error in our RG-improved one loop effective potential calculation of
$m_h$ which includes leading two-loop effects.)  The value of
$\Delta_{\rm EW}$ varies between $8.3-24$ along the model line so the
model is very natural with electroweak fine-tuning at the $12\%-4\%$
level.
\begin{table}\centering
\begin{tabular}{lc}
\hline
parameter & value \\
\hline
$m_0$      & 5000  \\
$m_{1/2}$   & 800  \\
$A_0$      & -8000  \\
$\tan\beta$&  10  \\
$\mu$      & 150 \\
$m_A$      &  1000  \\
\hline
$m_{\tg}$   & 2007.8   \\
$m_{\tu_L}$ &  5169.3  \\
$m_{\tu_R}$ &  5322.7  \\
$m_{\te_R}$ &  4808.0 \\
$m_{\tst_1}$&  1479.3  \\
$m_{\tst_2}$&  3650.1 \\
$m_{\tb_1}$ &  3678.3 \\
$m_{\tb_2}$ &  5049.3  \\
$m_{\ttau_1}$ & 4734.4  \\
$m_{\ttau_2}$ & 5079.7  \\
$m_{\tnu_{\tau}}$ & 5087.0 \\
$m_{\tw_2}$ & 691.3  \\
$m_{\tw_1}$ & 155.3 \\
$m_{\tz_4}$ & 702.2  \\ 
$m_{\tz_3}$ & 362.8  \\ 
$m_{\tz_2}$ & 158.2 \\ 
$m_{\tz_1}$ & 142.4 \\ 
$m_h$      & 124.4 \\ 
\hline
$\Omega_{\tz_1}^{std}h^2$ & 0.008  \\
$BF(b\to s\gamma)\times 10^4$ & 3.3  \\
$BF(B_s\to \mu^+\mu^-)\times 10^9$ & 3.8  \\
$\sigma^{SI}(\tz_1 p)$ (pb) & $4.3\times 10^{-9}$ \\
$\Delta_{\rm EW}$ & 10.3 \\
\hline
\end{tabular}
\caption{NUHM2 input parameters and masses in GeV units for a 
{\it radiatively-driven natural SUSY} benchmark points introduced in the text.
We take $m_t=173.2$~GeV}
\label{tab:bm}
\end{table}

\begin{figure}[tbp]
\begin{center}
\includegraphics[width=15cm,clip]{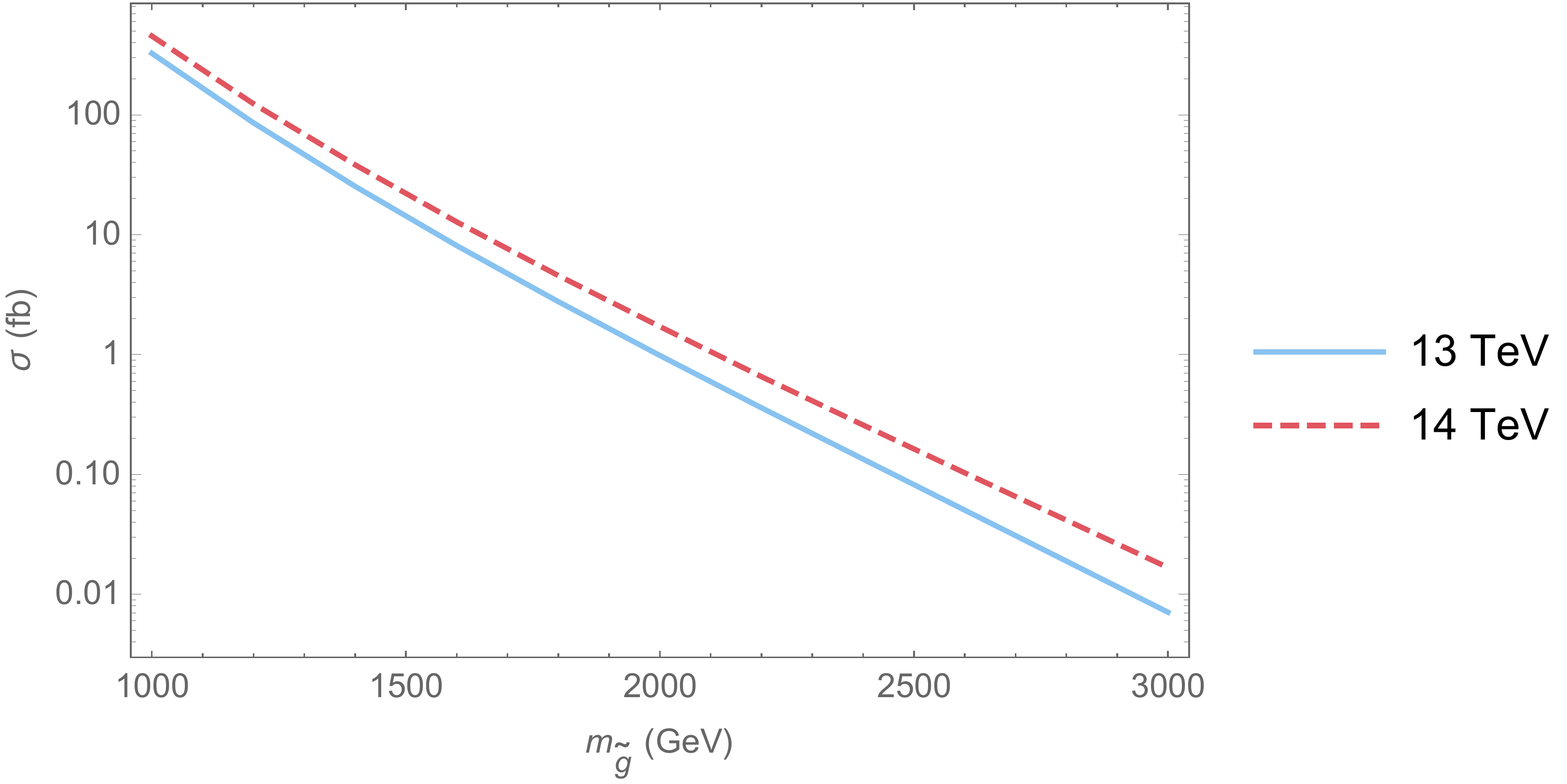}
\caption{Total NLO+NLL cross section for $pp\to\tg\tg X$ at LHC with
$\sqrt{s}=13$ and 14 TeV, versus $m_{\tg}$ for
$m_{\tq}\simeq 5$ TeV.
\label{fig:sigtot}}
\end{center}
\end{figure}

The cross section for $pp\to\tg\tg X$, calculated using
Prospino~\cite{Beenakker:1996ch} with {\sc
NLL-fast}~\cite{Beenakker:2011fu}, is shown in Fig.~\ref{fig:sigtot}
vs. $m_{\tg}$ for $m_{\tq}\simeq 5$ TeV and for $\sqrt{s}=13$ and 14
TeV.  For $m_{\tg}\sim 2$ TeV and $\sqrt{s}=14$ TeV -- the benchmark
case that we
adopt for devising our analysis cuts -- $\sigma (\tg\tg )\sim $ 1.7~fb;
the cross section drops to about $\sigma\sim 0.02$ fb for $m_{\tg}\sim
3$ TeV.

Once the gluinos are produced, all across the model line they decay
dominantly via the 2-body mode $\tg\to \tst_1 \bar{t}$ or $\tst_1^*
t$. For the benchmark point in Table~\ref{tab:bm}, the daughter top-squarks
rapidly decay via $\tst_1\to b\tw_1$ at $\sim 50\%$, $t\tz_1$ at $\sim
20\%$, $t\tz_2$ at $\sim 24\%$ and $t\tz_3$ at $\sim 6\%$. Stop decays
into $b\tw_2$ and $t\tz_4$ are suppressed since in our model with stop
soft masses unified at $m_0$ at the GUT scale, then the $\tst_1$ is
mainly a right-stop eigenstate with suppressed decays to winos. The
stop branching fractions vary hardly at all as $m_{1/2}$ varies along the
model line.  The higgsino-like $\tz_1$ state is expected to comprise a
portion of the dark matter in the universe (the remaining portion might
consist of, {\it e.g.}, axions~\cite{axdm}) while the higgsino-like $\tz_2$
and $\tw_1$ decay via 3-body modes to rather soft visible debris because
the mass gaps $m_{\tz_2}-m_{\tz_1}$ and $m_{\tw_1}-m_{\tz_1}$
are typically only 10-20 GeV and hence essentially
invisible for the purposes of this paper.

Putting together production and decay processes, gluino pair
production final states consist of $t\bar{t}t\bar{t}+\eslt$,
$t\bar{t}t\bar{b}+\eslt$ and $t\bar{t}b\bar{b}+\eslt$ parton
configurations. In the case where $\tz_2$ is produced via the gluino
cascade decays, then the boosted decay products from
$\tz_2\to\ell^+\ell^-\tz_1$ decay may display an invariant mass edge
$m(\ell^+\ell^- )<m_{\tz_2}-m_{\tz_1}\sim 10-20$ GeV. The existence of
such an edge in gluino cascade decay events containing an OS/SF dilepton
pair would herald the presence of light higgsinos~\cite{lhc,baris} though
the cross sections for these events are very small. In this paper, our
focus will be on the observation of the signal and prospects 
for gluino mass reconstruction using the inclusive  sample with
$t\bar{t}t\bar{t}+\eslt$,
$t\bar{t}t\bar{b}+\eslt$ and $t\bar{t}b\bar{b}+\eslt$ final states,
with no attention to how the final state higgsinos (which are produced in the bulk
of the cascade decays) decay.

\section{Event generation}
\label{sec:evgen}

We employ two procedures for event generation, one using {\sc
Isajet}~7.85~\cite{isajet}, which we refer to as our ``Isajet''
simulation and one using {\sc MadGraph}~2.3.3~\cite{madgraph} interfaced
to {\sc PYTHIA} 6.4.14~\cite{pythia} with detector simulation by {\sc
Delphes}~3.3.0~\cite{delphes}, which we refer to as our ``MadGraph''
simulation.

\subsection{Isajet Simulation}

Our Isajet simulation includes detector simulation by the Isajet toy
detector, with calorimeter cell size $\Delta \eta \times \Delta \phi =
0.05 \times 0.05$ and $-5 < \eta < 5$.  The HCAL energy resolution is
taken to be $80\%/\sqrt{E} + 3\%$ for $|\eta| < 2.6$ and $100\%/\sqrt{E}
+ 5\%$ for $|\eta| > 2.6$, where the plus denotes combination in
quadrature.  The ECAL energy resolution is assumed to be $3\%/\sqrt{E} +
0.5\%$.  We use a UA1-like jet finding algorithm with jet cone size $R =
0.4$ and require that $E_T(jet) > 50$ GeV and $|\eta(jet)| < 3.0$.
Leptons are considered isolated if they have $p_T(e~or~\mu) > 20$ GeV
and $|\eta| < 2.5$ with visible activity within a cone of $\Delta R <
0.2$ of $\sum E_T^{cells} < 5$ GeV.  The strict isolation criterion
helps reduce multi-lepton backgrounds from heavy quark ($c\bar{c}$ and
$b\bar{b}$) production.

We identify a hadronic cluster with $E_T > 50$ GeV and $|\eta(jet)| <
1.5$ as a $b$-jet if it contains a $B$ hadron with $p_T(B) > 15$ GeV and
$|\eta(B)| < 3$ within a cone of $\Delta R < 0.5$ around the jet axis.
We adopt a $b$-jet tagging efficiency of $60\%$ and assume that light
quark and gluon jets can be mistagged as $b$-jets with a a probability
of $1/150$ for $E_T < 100$ GeV, $1/50$ for $E_T > 250$ GeV and a linear
interpolation for $100$ GeV $ < E_T < 250$ GeV.\footnote{These values are based
on ATLAS studies of $b$-tagging efficiencies and rejection factors in
$t\bar{t}H$ and $WH$ production processes~\cite{atlasb}.} We refer to these
values as our ``Isajet'' parameterization of $b$-tagging efficiencies.

\subsection{MadGraph Simulation}

In our MadGraph simulation, the events are showered and hadronized using
the default MadGraph/PYTHIA interface with default parameters.
Detector simulation is performed by Delphes using the default
Delphes~3.3.0 ``CMS'' parameter card with several changes, which we
enumerate here.
\begin{enumerate}

\item We set the HCAL and ECAL resolution formulae to be those that we
have used in our Isajet simulation.

\item We turn off the jet energy scale correction.

\item We use an anti-$k_T$ jet algorithm~\cite{Cacciari:2008gp} with $R
= 0.4$ rather than the default $R = 0.5$ for jet finding in Delphes
(which is implemented via {\sc FastJet}~\cite{Cacciari:2011ma}).  As in
our Isajet simulation, we only consider jets with $E_T(jet) > 50$ GeV
and $|\eta(jet)| < 3.0$ in our analysis.  The choice of $R = 0.4$ in the
jet algorithm is made both to make our MadGraph simulation conform to
our Isjaet simulation and to allow comparison with CMS $b$-tagging
efficiencies~\cite{CMS:2016kkf}: see Table~\ref{tab:bcomp} below. 

\item We write our own jet flavor association module based on the
``ghost hadron'' procedure~\cite{Cacciari:2007fd}, which allows decayed
hadrons to be unambiguously assigned to jets.  With this functionality
we identify a jet with $|\eta| < 1.5$ as a $b$-jet if it contains a $B$
hadron (in which the $b$ quark decays at the next step of the decay)
with $|\eta| < 3.0$ and $p_T > 15$ GeV.  These values are in accordance
with our Isajet simulation.

\item We turn off tau tagging, as we do not use the tagging of hadronic
taus in our analyses.  Sometimes Delphes will wrongly tag a true $b$-jet as a
tau, if the $B$ hadron in the jet decays to a tau.  As we are trying to
perform a cross section measurement in a regime where the overall signal
cross section is small, we did not want to ``lose'' these $b$-jets.

\end{enumerate}

\subsection{Processes Simulated}

Our Isajet simulation was used to generate the signal from gluino pair
production at our benchmark point, as well as for other parameter points
along our model line.  We also used our Isajet simulation to simulate
backgrounds from $t\bar{t}$, $W$ + jets, $Z$ + jets, $WW$, $WZ$, and
$ZZ$ production.  The $W$ + jets and $Z$ + jets backgrounds use exact
matrix elements for one parton emission, but rely on the parton shower
for subsequent emissions.  In addition, we have generated background
events with our Isajet simulation procedure for QCD jet production
(jet-types including $g$, $u$, $d$, $s$, $c$, and $b$ quarks) over five
$p_T$ ranges, as shown in Table II of Ref.~\cite{gabe}.  Additional jets
are generated via parton showering from the initial and final hard
scattering subprocesses.

Our MadGraph simulation was used to generate the signal from gluino pair
production at our benchmark point, as well as for other parameter points
along our model line, and also to generate backgrounds from
$t\bar{t}$, $t\bar{t}b\bar{b}$, $b\bar{b}Z$, and $t\bar{t}t\bar{t}$
production.  To avoid the double counting that would ensue from
simulating $t\bar{t}$ as well as $t\bar{t}b\bar{b}$, we veto events with
more than two truth $b$-jets in our $t\bar{t}$ sample.

In simulating $t\bar{t}$, $t\bar{t}b\bar{b}$, and $b\bar{b}Z$ with
MadGraph, we generate events in various bins of generator-level $\eslt$.
The use of weighted events from this procedure gives us sensitivity to
the high tail of the $\eslt$ distribution for these background
processes.  This sensitivity is essential for determining the rates that
remain from background processes after the very hard $\eslt$ cuts,
described in the next section, that
we use to isolate the signal.

In our MadGraph simulation, we normalize the overall cross section for
our signal to NLL values obtained from {\sc
NLL-fast}~\cite{Beenakker:2011fu}.  For $t\bar{t}$ we used an overall
cross section of $953.6$ pb, following Ref.~\cite{Czakon:2013goa}.  As
MadGraph chooses the scale dynamically event-by-event, we follow
Ref.~\cite{Bredenstein:2010rs} and use a K-factor of $1.3$ for our
$t\bar{t}b\bar{b}$ backgrounds; the authors of this work find larger
K-factors when a dynamic scale choice is not employed~\cite{bred2}.  For
the evaluation of the background from $b\bar{b}Z$ production we use a
K-factor of $1.5$, following Ref.~\cite{Cordero:2009kv}, while for the
$t\bar{t}t\bar{t}$ backgrounds we use a K-factor of $1.27$, following
Ref.~\cite{Bevilacqua:2012em}.

We found very similar results when using signal events from our Isajet
simulation procedure as when using signal events from our MadGraph
simulation procedure.  We found significantly more $t\bar{t}$ events
with high values of missing $\eslt$ from our MadGraph simulation
procedure than we did from our Isjaet procedure, presumably due to
differences in showering algorithms.  To be conservative, we use the
larger $t\bar{t}$ backgrounds generated from MadGraph in our analyses.
The hard $\eslt$ cuts described below together with the requirement of
at least two tagged $b$-jets, very efficiently remove the backgrounds 
from $W,Z$+ jets and from $VV$ production simulated with Isajet.
In the interest of presenting a clear and concise description of our analysis,
we will not include these backgrounds in the figures and tables in the
remainder of this work.  For consistency with the most relevant SM backgrounds
from $t\bar{t}$, $Zb\bar{b}$, $t\bar{t}b\bar{b}$ and $t\bar{t}t\bar{t}$
production, we likewise utilize our signal samples generated using the
MadGraph simulation procedure.

\section{Gluino Event Selection}
\label{sec:cuts}
To separate the gluino events from SM backgrounds,  
we begin by applying a set of pre-cuts to our event samples, which we
call {\bf C1} (for ``cut set 1'').  These are very similar to a set of cuts
found in the literature~\cite{gabe, frank}.  However, since our focus is
on the signal from very heavy gluinos ($m_{\tg} \ge 1.6$~TeV), we have
raised the cut on jet $p_T$ to $100$ GeV from $50$ GeV and included a cut
on the transverse mass of the lepton and $\eslt$  
in events with only one isolated lepton 
(to reduce backgrounds from events with $W$ bosons).
\\
\\
\textbf{C1 Cuts:}
\bea
\eslt & >& {\rm max}(100\ {\rm GeV},0.2 M_{eff}),\nonumber\\
n(jets) &\ge & 4,\nonumber \\
E_T(j_1,j_2,j_3,j_4)& > & 100\ {\rm GeV},\\
S_T &>&0.2 , \nonumber \\
m_T(\ell, \eslt) & > & 150\ {\rm GeV,~if~} n_{lep} = 1. \nonumber
\label{c1cuts}
\eea
Here, $M_{eff}$ is defined as in Hinchliffe {\it et al.}~\cite{frank} as
$M_{eff}=\eslt +E_T(j_1)+E_T(j_2)+E_T(j_3)+E_T(j_4)$, where $j_1-j_4$ refer to
the four highest $E_T$ jets ordered from highest to lowest $E_T$,
$\eslt$ is missing transverse energy, $S_T$ is transverse
sphericity\footnote{Sphericity is defined, {\it e.g.}, in {\it Collider
Physics}, V. Barger and R. J. N. Phillips (Addison Wesley, 1987) Here,
we restrict its construction to using only transverse quantities, as is
appropriate for a hadron collider.}, and $m_T$ is the transverse mass of
the lepton and the $\eslt$.
\begin{figure}[tbp]
\begin{center}
\includegraphics[width=\lfig,clip]{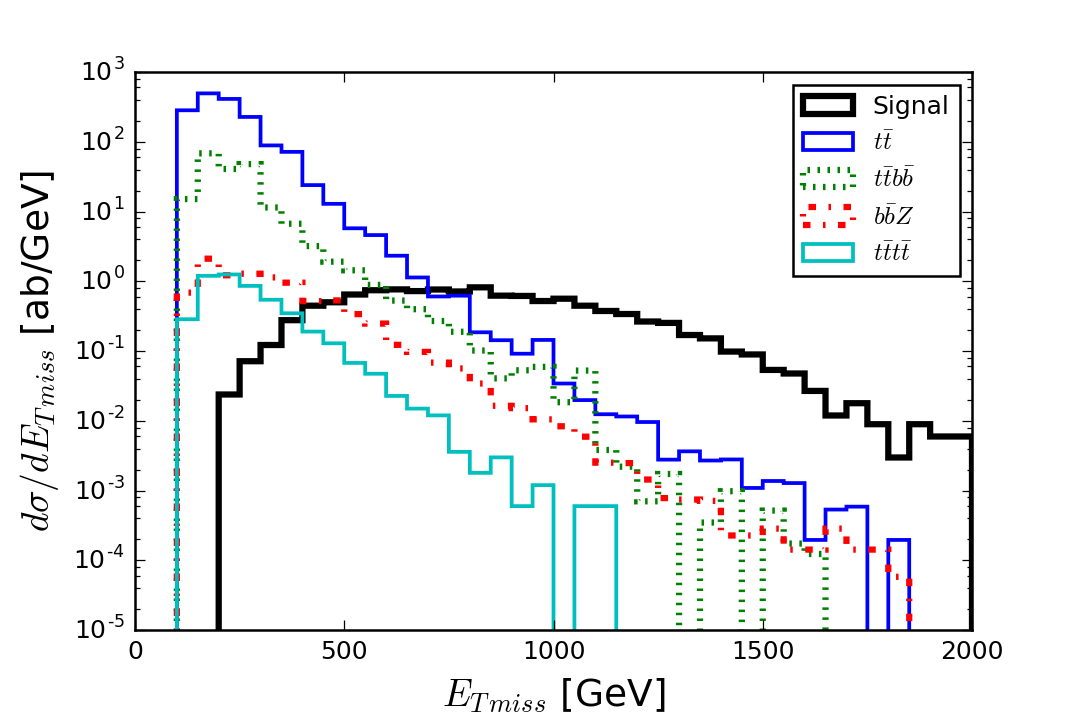}
\caption{
Distribution of $\eslt$ after C1 cuts~(\ref{c1cuts}) with the requirement of
two $b$-tagged jets for the gluino pair production signal, as well as the most
relevant backgrounds ($t\bar{t}$, $t\bar{t}b\bar{b}$, $b\bar{b}Z$, and $t\bar{t}t\bar{t}$).
\label{fig:fig2}}
\end{center}
\end{figure}

Since the signal naturally contains a high multiplicity of hard $b$-partons
from the decay of the gluinos because third generation squarks tend to
be lighter than other squarks, in addition to the basic {\bf C1} cuts,
we also require the presence of two tagged $b$-jets,
\\
\\
\textbf{$b$-jet multiplicity Cut:}
\bea
n_b & \ge & 2.
\label{bjet_cut}
\eea 
using the ``Isajet'' parameterization of $b$-tagging efficiencies
and light jet mistagging.

Even after these cuts, we must still contend with sizable backgrounds,
as can be seen from Fig.~\ref{fig:fig2} where we show the $\eslt$
distribution from the $t\bar{t}$, $t\bar{t}b\bar{b}$, $b\bar{b}Z$, and
$t\bar{t}t\bar{t}$ backgrounds, as well as from the gluino pair
production for the benchmark point in Table~\ref{tab:bm}.
We see that the
backgrounds fall more quickly with $\eslt$ than the signal leading us
to impose a $\eslt$ cut,
\\
\\
\textbf{$\eslt$ Cut:}
\bea
\eslt & > & 750\ {\rm GeV}.
\label{missing_et_cut}
\eea
\begin{figure}[tbp]
\begin{center}
\includegraphics[width=\lfig,clip]{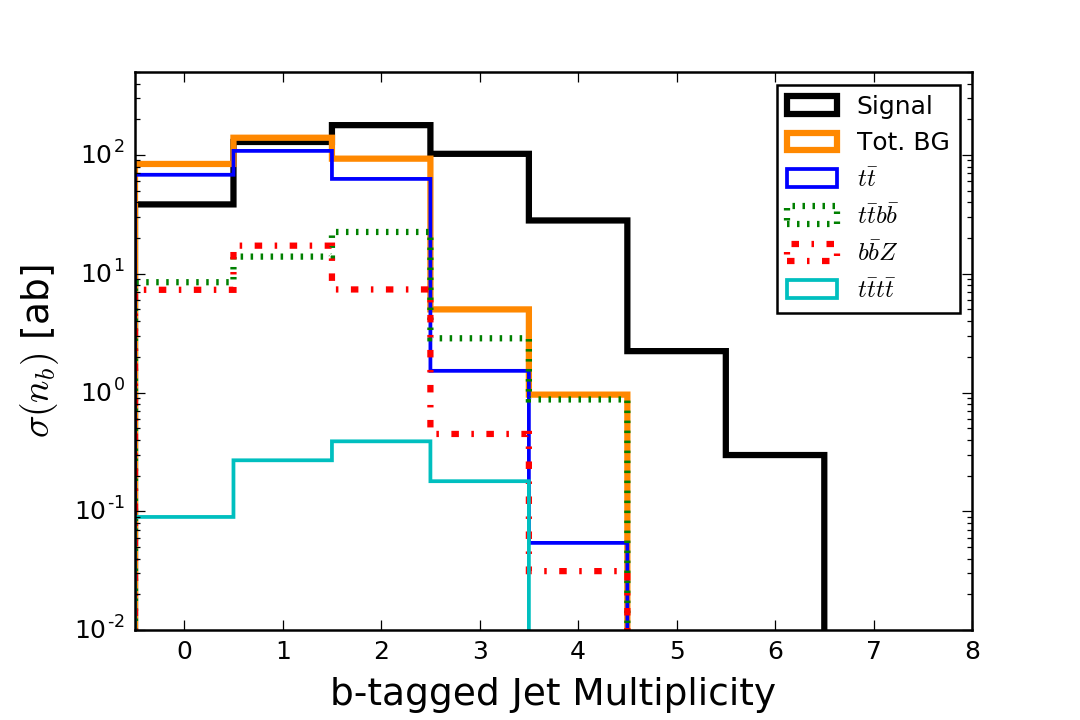}
\caption{
The number of $b$-tagged jets, using the Isajet parameterization of the
$b$-tagging efficiency, after C1 cuts~(\ref{c1cuts}) and the requirement
that $\eslt > 750$ GeV.
\label{fig:fig3}}
\end{center}
\end{figure}
After this cut, we are left with comparable backgrounds from $t\bar{t}$
and $t\bar{t}b\bar{b}$ production with a somewhat smaller contribution
from $b\bar{b}Z$ production. The $t\bar{t}t\bar{t}$ background rate 
is much smaller.

Once we have made the $\eslt$ cut~(\ref{missing_et_cut}), we examine the
distribution of the multiplicity of $b$-tagged jets, with the goal of
further improving the signal to background ratio. 
%
This distribution is shown in Fig.~\ref{fig:fig3}.  This figure suggests
two roads to selection criteria that will leave a robust signal and
negligible backgrounds.  Obviously, we can require three $b$-tags, which
decimates the backgrounds (especially $t\bar{t}$) at the cost of some
signal. Our goal is to devise a strategy that will allow mass
measurements even with integrated luminosities of 100-200~fb$^{-1}$ that
will be available by the end of the 2018 LHC shutdown for which
significant loss of event rate rapidly becomes a problem. With this in
mind, we also examine the possibility that we can only require two
$b$-tags.  While this saves some signal, we clearly need to impose
additional cuts to obtain a clean signal sample.  We pursue both of
these approaches: the larger cross section from the ``$2b$'' analysis
will certainly be useful in early LHC running, but the greater reduction of
backgrounds provided by the ``$3b$'' analysis would be expected to yield
cleaner data samples at the
high luminosity LHC. 

\begin{figure}[tbp]
\begin{center}
\includegraphics[width=\sfig,clip]{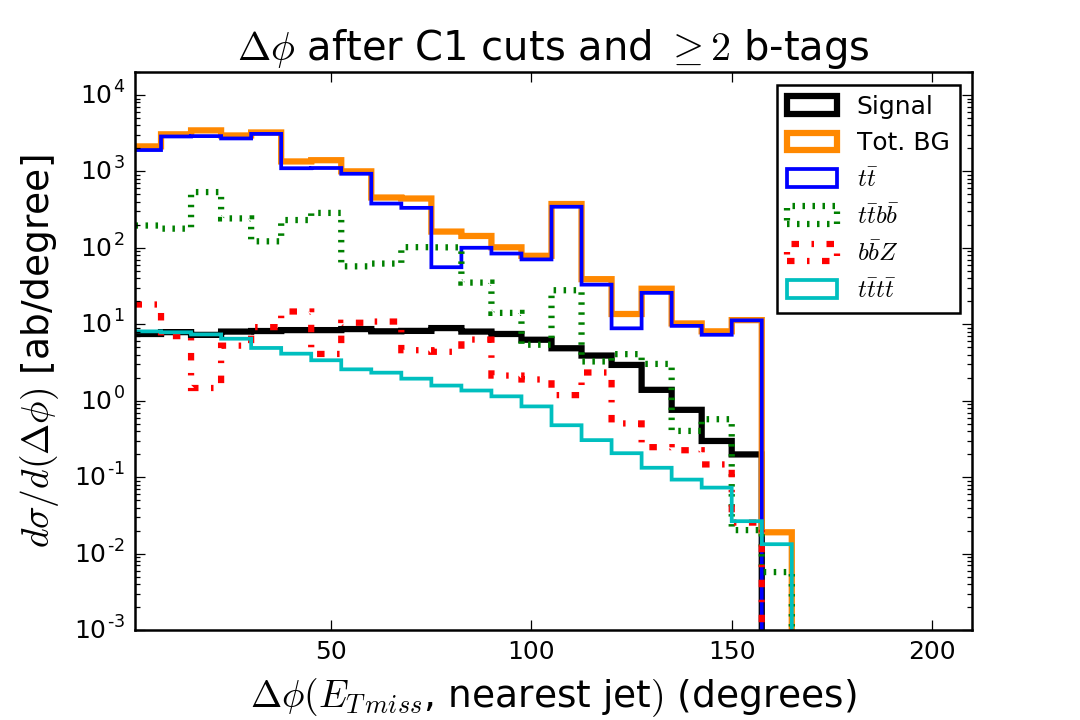}
\includegraphics[width=\sfig,clip]{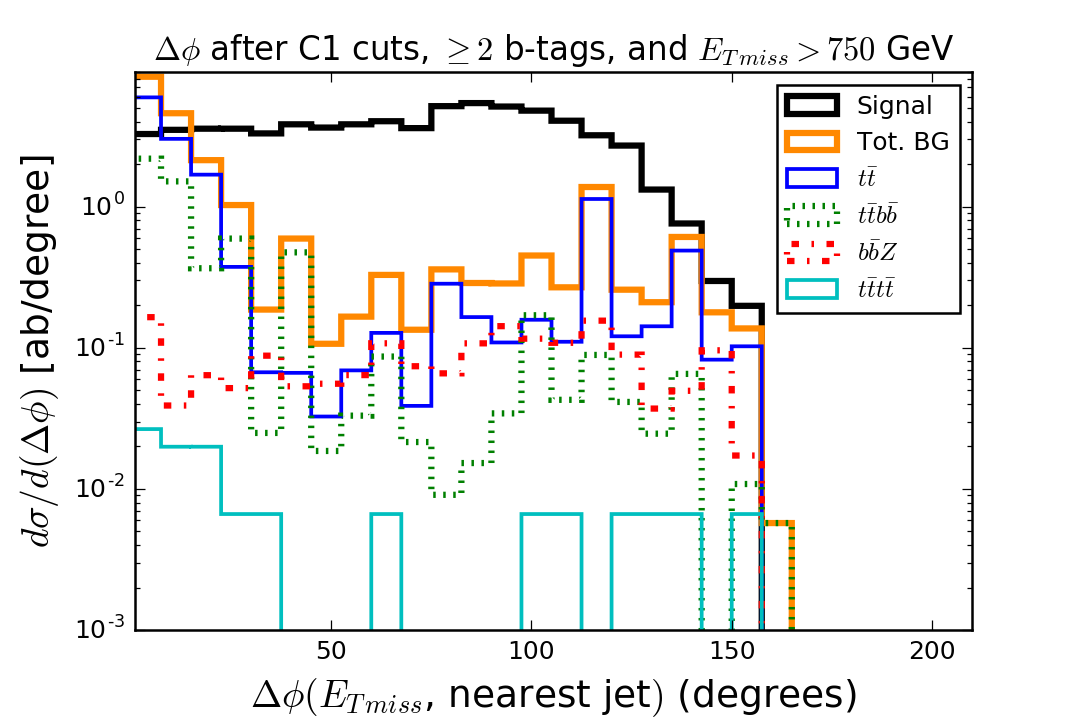}
\caption{
The distribution of $\Delta \phi(\eslt,\ {\rm nearest~jet})$.
Explicitly this quantity is the minimum angle between the $\eslt$ vector and
the transverse momentum of one of the leading four jets.  This
quantity is shown (left) after C1 cuts~(\ref{c1cuts}) with the requirement
of two $b$-tagged jets and (right) after these cuts, and a cut of $\eslt > 750$ GeV.
\label{fig:fig4}}
\end{center}
\end{figure}

To further clean up the $n_b\ge 2$ signal sample, we first note that that
the bulk of the background comes from $t\bar{t}$ production. It is
reasonable to expect that $t\bar{t}$ production 
leads to $\eslt > 750$~GeV only if a
semi-leptonically decaying top is produced with a very high transverse
momentum, with the daughter neutrino ``thrown forward'' in the top rest
frame, while the other top decays hadronically (so the $\eslt$ is not
cancelled). In this case, the $b$-jet from the decay of the
semi-leptonically decaying top would tend to be collimated with the
neutrino; {\it i.e.}, to the direction of $\eslt$. We do not expect such a
correlation in the signal since the heavy gluinos need not be
particularly boosted to yield $\eslt > 750$~GeV.
%
This motivated us to examine the distribution of the minimum value of
$\Delta \phi$, the angle between the transverse momenta of a jet and the
$\eslt$ vector, for each of the four leading jets.  We show this
distribution in Fig.~\ref{fig:fig4}, after the {\bf C1} and the two
tagged $b$-jet cuts, both with (right frame) and without (left frame)
the $\eslt > 750$ GeV cut.  Without this hard $\eslt$ cut, we see that
the distribution of $\Delta\phi$ is very slowly falling for the
$t\bar{t}$ background, and roughly flat for the signal as for the other
backgrounds, until all the distributions cut-off at about
$150^\circ$. The expected peaking of the $t\bar{t}$ background at low
values of $\Delta\phi$ is, however, clearly visible in the right frame,
while the signal is quite flat. The next-largest background from
$t\bar{t}b\bar{b}$ also shows a similar peaking (for the same reason) at
low $\Delta\phi$ values. We are thus led to impose the cut,\\ \\
\textbf{$\Delta \phi$ Cut:} \bea \Delta \phi (\eslt,
\rm{nearest~of~four~leading~jets}) & > & 30^\circ,
\label{delta_phi_cut}
\eea
which greatly diminishes the dominant backgrounds in the two tagged
$b$-jet channel with only a very modest loss of signal. Indeed, because
the signal-to-background ratio is so vastly improved with only a slight
reduction of the signal, we have retained this cut in both our 2$b$ and
$3b$ analyses.

\begin{figure}[tbp]
\begin{center}
\includegraphics[width=\sfig,clip]{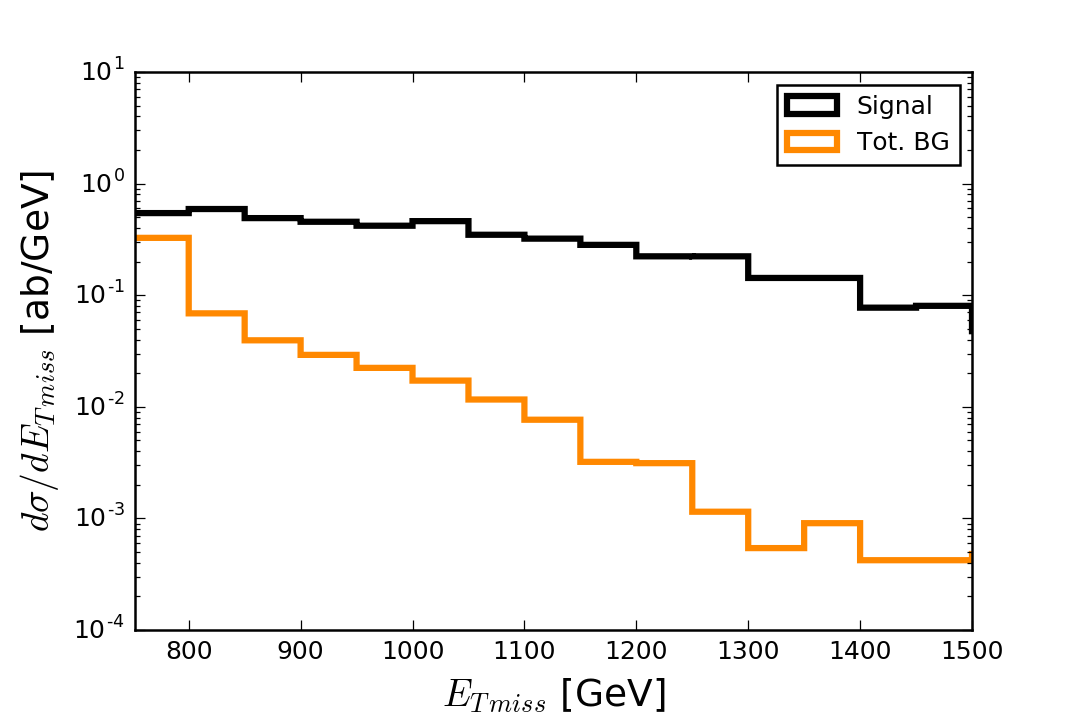}
\includegraphics[width=\sfig,clip]{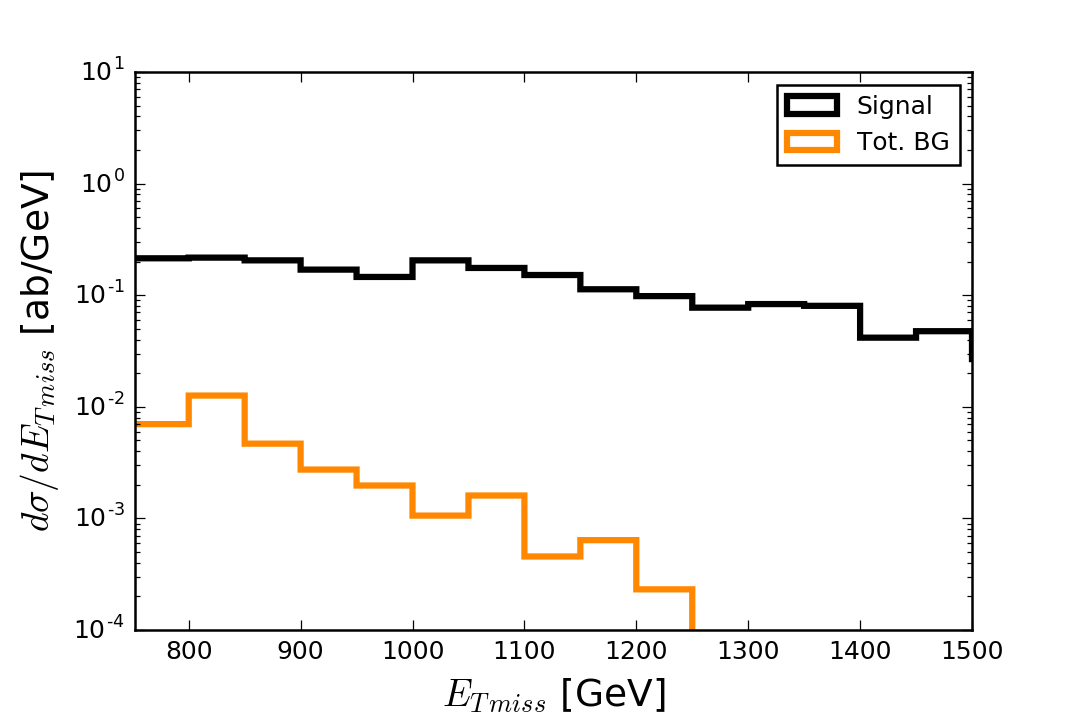}
\caption{ The distribution of $\eslt$ after {\bf C1} cuts~(\ref{c1cuts}), the
$\eslt > 750$ cut~(\ref{missing_et_cut}), and the $\Delta \phi >
30^\circ$ cut~(\ref{delta_phi_cut}) with the additional requirement of
(left) at least two $b$-tagged jets (right) at least three $b$-tagged
jets.  The background distribution represents the sum of the
contributions from the $t\bar{t}$, $t\bar{t}b\bar{b}$, $b\bar{b}Z$, and
$t\bar{t}t\bar{t}$ backgrounds.
\label{fig:fig5}}
\end{center}
\end{figure}

Having made this cut, we return to the $\eslt$ distribution, to see
whether further optimization might be possible.  Toward this end, we
show the distribution after the {\bf C1} cuts (\ref{c1cuts}), the $\eslt >
750$ GeV cut (\ref{missing_et_cut}), and the $\Delta \phi > 30^\circ$
cut (\ref{delta_phi_cut}) in Fig.~\ref{fig:fig5}, requiring at least two
$b$-tagged jets (left panel) or three $b$-tagged jets (right panel).  We
see that an additional cut on $\eslt$ will be helpful in the $2b$
analysis, but not as helpful in the $3b$ analysis.  Therefore, our final
cut choices are: \\ \\ \textbf{$2b$ Analysis:} \bea {\rm C1~cuts},\nonumber \\ n_b
& \geq & 2,\nonumber  \\ \Delta \phi (\eslt, \rm{nearest~of~four~leading~jets}) &
> & 30^\circ, \label{2b_analysis} \\ \eslt & > & 900\ {\rm GeV},\nonumber
\eea
\begin{center}
and
\end{center}
\textbf{$3b$ Analysis:}
\bea
{\rm C1~cuts}, \nonumber \\
n_b & \geq & 3,\nonumber  \\
\Delta \phi (\eslt, \rm{nearest~of~four~leading~jets}) & > & 30^\circ, \label{3b_analysis} \\
\eslt & > & 750\ {\rm GeV},\nonumber
\eea
The cross section including acceptance after each of the cuts, for the
signal benchmark point, as well as for the sum of the $t\bar{t}$,
$t\bar{t}b\bar{b}$, $b\bar{b}Z$ and $t\bar{t}t\bar{t}$ backgrounds, is
given in Table~\ref{tab:cut_flow}, for both the $2b$ and the $3b$
analyses. 
\begin{table}\centering
\begin{tabular}{ | l | c | c | c | c | } \hline
Cut & $2b$ Sig.          & $2b$ BG & $3b$ Sig. & $3b$ BG \\ \hline \hline
C1                       & $872$ & $4.99 \times 10^5$ & $872$ & $4.99 \times 10^5$ \\ \hline
$\eslt > 750$ GeV        & $479$ & $323$              & $479$ & $323$              \\ \hline
$b$-tagging              & $311$ & $99.1$             & $133$ & $5.98$             \\ \hline
$\Delta \phi > 30^\circ$ & $249$ & $26.8$             & $105$ & $1.65$             \\ \hline
Final $\eslt$ cut        & $167$ & $5.02$             & $105$ & $1.65$             \\ \hline
\end{tabular}
\caption{ Cross section times acceptance in attobarns (1000 ab$=$ 1 fb) 
after various cuts
are applied.  The ``$b$-tagging'' cut refers to the requirement of $\geq
2$ $b$-tagged jets in the $2b$ analysis and $\geq 3$ $b$-tagged jets in
the $3b$ analysis.  For the $2b$ analysis, the ``final $\eslt$ cut''
refers to the additional requirement that $\eslt > 900$ GeV; there is no
additional cut in the $3b$ analysis.  }
\label{tab:cut_flow}
\end{table}

\begin{figure}[tbp]
\begin{center}
\includegraphics[width=\sfig,clip]{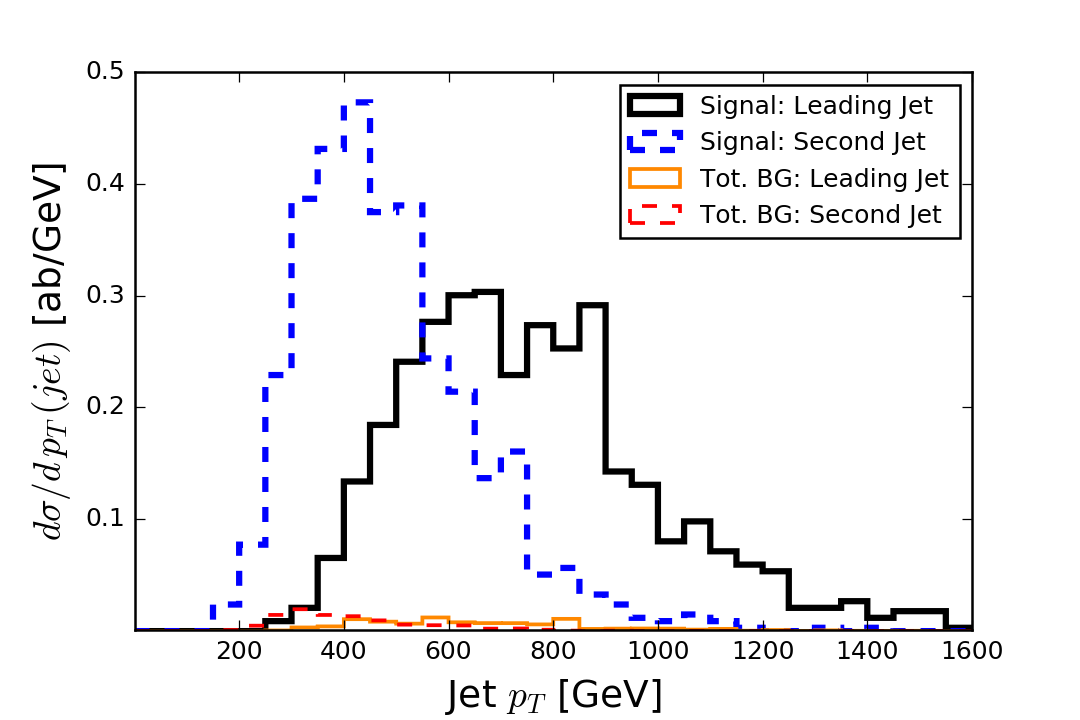}
\includegraphics[width=\sfig,clip]{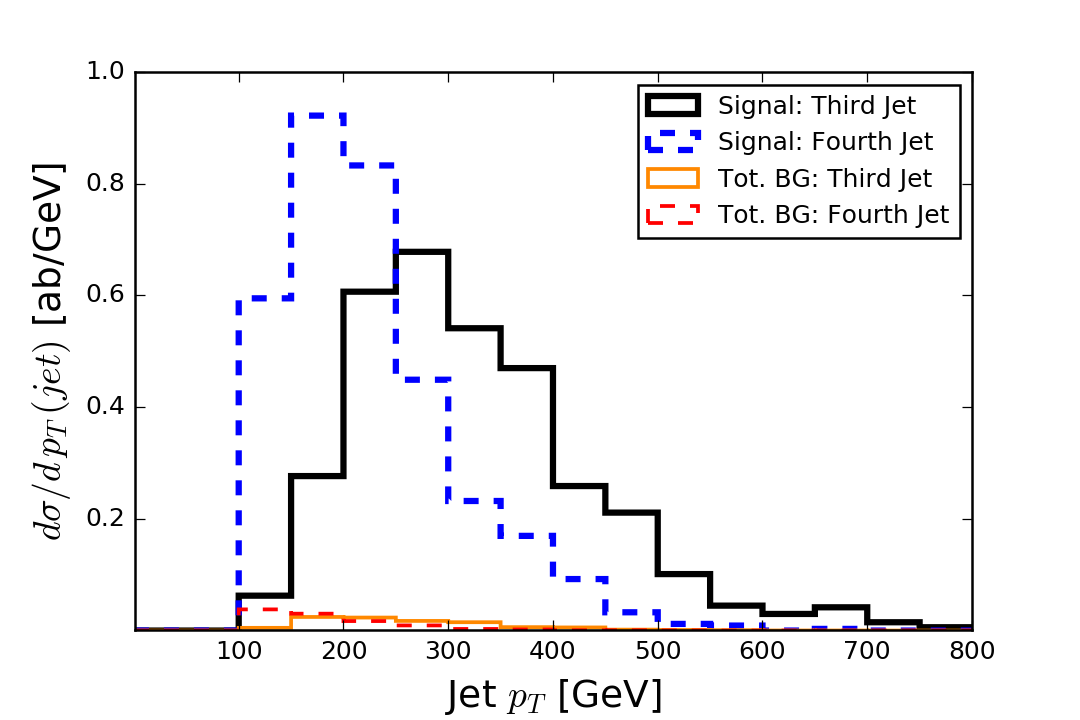}
\caption{
Transverse momenta of the leading jet and second-leading jet in $p_T$
(left) and for the third and fourth-leading jets (right) for signal and background
events after $2b$ analysis cuts.
The distribution of these quantities after $3b$ analysis cuts is similar.
\label{fig:fig6}}
\end{center}
\end{figure}

\subsection{Gluino Event Characteristics}

Now that we have finalized our analysis cuts, we display the
characteristic features of gluino signal events satisfying our selection
criteria for our natural SUSY benchmark point with $m_{\tg}\simeq 2$~TeV
and $m_{\tst_1} \sim 1500$~TeV.  Figure~\ref{fig:fig6} shows the
transverse energy distribution of the four hardest jets from the
two-tagged $b$-jet signal as well as from the backgrounds, after the cut
set (\ref{2b_analysis}). We see that the two hardest jets typically have
$E_T\sim 700$~GeV and 400~GeV, respectively, while the third and fourth
jet $E_T$ distributions peak just below 300~GeV and 200~GeV. The
distributions for the signal with three tagged $b$-jets are very similar
and not shown for brevity. While the actual peak positions in the
distributions depend on the gluino and stop masses, the fact that the
events contain four hard jets is rather generic. We also see that the SM
background after these cuts is negligibly small, and that we do indeed
have a pure sample of gluino events.

\begin{figure}[tbp]
\begin{center}
\includegraphics[width=\lfig,clip]{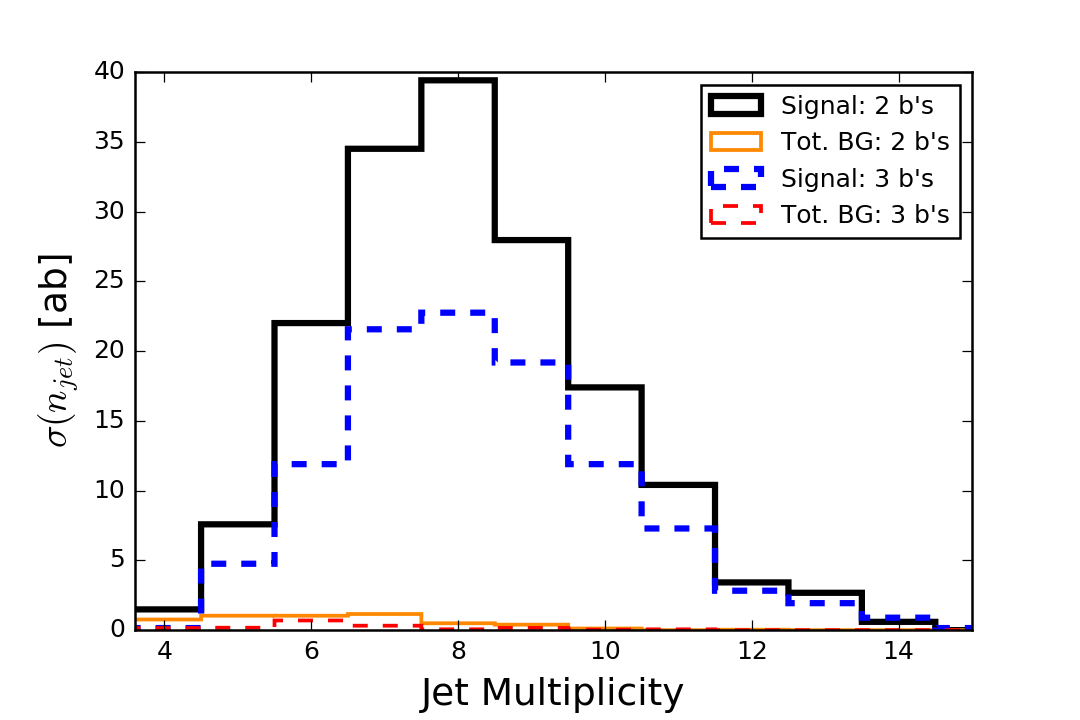}
\caption{
Jet multiplicity for signal and background events satisfying our $2b$ and $3b$
analysis cuts.  Recall that we require jets to have $p_T > 50$ GeV and $|\eta| < 3.0$.
\label{fig:fig7}}
\end{center}
\end{figure}

In Fig.~\ref{fig:fig7}, we show the jet multiplicity for the benchmark
point signal and background events after our selection cuts for both the
two tagged $b$-jet (solid) and the three-tagged $b$-jet (dashed)
samples. Recall that jets are defined to be hadronic clusters with $E_T>
50$~GeV. We see that the signal indeed has very high jet multiplicity
relative to the background. Since the exact jet multiplicity may be
sensitive to details of jet definition, and because our simulation of
the background with very high jet multiplicity is less reliable due to
the use of the shower approximation rather than exact matrix
elements, we have {\it not used} jet multiplicity cuts to further
enhance the signal over background. (Note: the sum of cross-sections 
above a minimum jet-multiplicity, as implemented in the {\bf C1} cuts, 
is not expected to depend much 
on the implementation of the jet multiplicity cut.)

\begin{figure}[tbp]
\includegraphics[width=\sfig,clip]{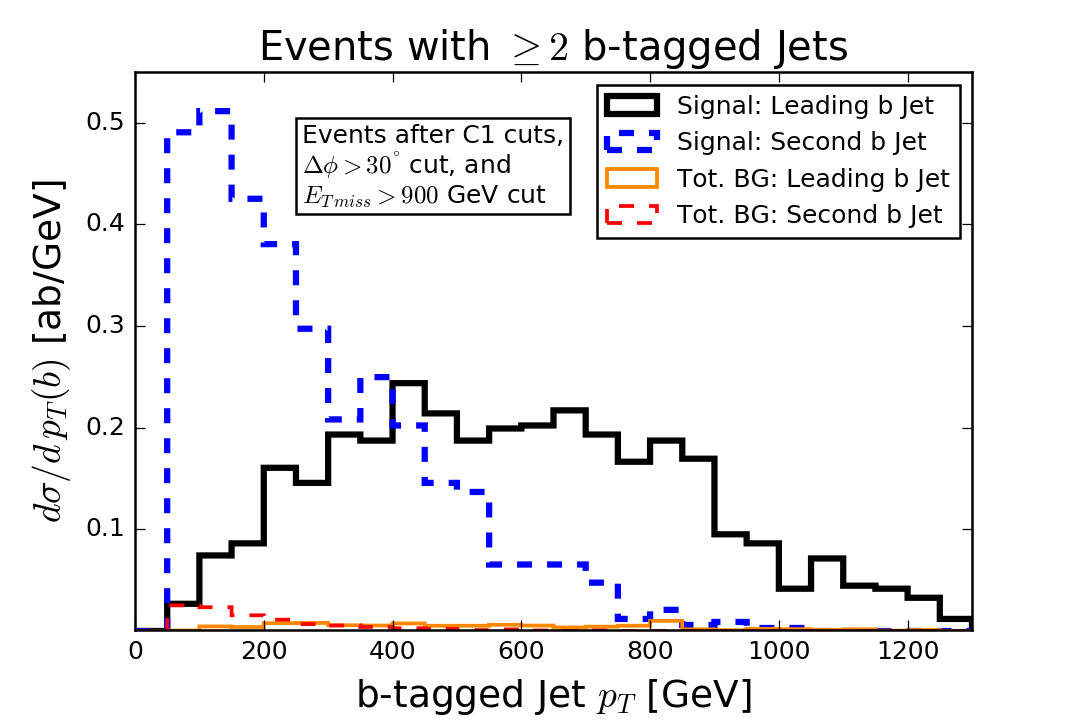}
\includegraphics[width=\sfig,clip]{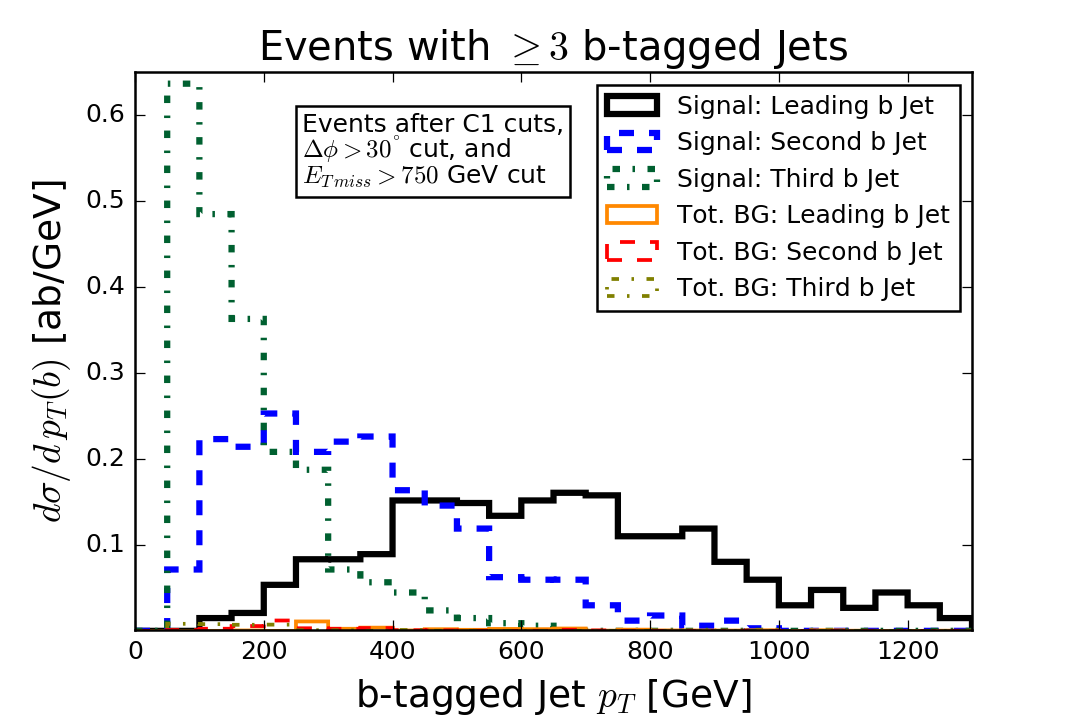}
\caption{
The distribution of the transverse momenta of the leading
and second leading $b$-tagged jet after (left) the $2b$ analysis cuts and
(right) the $3b$ analysis cuts.
\label{fig:fig8}}
\end{figure}

In Fig.~\ref{fig:fig8} we show the transverse momentum of $b$-tagged
jets in signal and background events satisfying the final cuts for $\ge
2$ tagged $b$-jet events (left frame) and for $\ge 3$ tagged $b$-jet
events (right frame). We see that the hardest $b$-jet $E_T$
ranges up to $\sim 1$ TeV, while the second $b$-jet, for the most
part, has $E_T \sim 100-600$~GeV. Again, we stress that the $b$-jet
spectrum shape will be somewhat sensitive to the gluino-stop as well as
stop-higgsino mass differences, but the hardness of the $b$-jets is
quite general. We expect that the $b$-jets would remain hard (though the
$E_T$ distributions would have different shapes) even in the case when
the stop is heavier than the gluino, and the gluino instead dominantly
decays via the three body modes, $\tg \to t\bar{t}\tz_{1,2}$ and $\tg
\to tb\tw_1$.

Before turning to a discussion of our results for the mass reach and of
the feasibility of the extraction of $m_{\tg}$ using the very pure
sample of signal events, we address the sensitivity of our cross section
calculations to the Isajet $b$-tagging efficiency and purity algorithm
that we have used. This algorithm was based on early ATLAS studies
~\cite{atlasb} of $WH$ and $t\bar{t}H$ processes where the transverse
momentum of the $b$-jets is limited to several hundred GeV. More
recently, the CMS Collaboration~\cite{CMS:2016kkf} has provided loose,
medium and tight $b$-tagging algorithms with corresponding charm and
light parton mis-tags whose validity extends out to a TeV. We show a
comparison of the SUSY signal rate for our SUSY benchmark point for the
sample with at least two/three tagged $b$-jets after the selection cuts
(\ref{2b_analysis})/(\ref{3b_analysis}) in Table \ref{tab:bcomp}.  We
illustrate results for the medium and tight algorithms in
Ref.~\cite{CMS:2016kkf}. Also shown, in parenthesis are the corresponding
signal-to-background ratios, after these cuts.  We see that the cross
sections for the Isajet parametrization of the $b$-tagging efficiency,
as well as the corresponding values of $S/B$ lie between those obtained
using the medium and tight algorithms in the recent CMS study. Although
it is difficult to project just how well $b$-tagging will perform in the
high luminosity environment, we are encouraged to see that our simple
algorithm gives comparable answers to those obtained using the more
recent tagging algorithms in Ref.~\cite{CMS:2016kkf} even though we have
very hard $b$-jets in the signal.
\begin{table}\centering
\begin{tabular}{ | l | c | c | c | } \hline
 & Isajet & CMS Medium & CMS Tight \\ \hline \hline $\ge 2$ tagged $b$ jets,
$\eslt > 900$~GeV & 167 \ (33) & 207 \ (26) & 121 \ (41) \\ \hline $\ge
3$ tagged $b$ jets, $\eslt > 750$~GeV & 105 \ (66) & 182 \ (50) & 61.1 \ (81)
\\ \hline
\end{tabular}
\caption{The LHC signal cross section in ab for our SUSY benchmark point
  for $\ge 2$ tagged $b$-jet events, and for $\ge 3$ tagged $b$-jet
  events after all the analysis cuts in (\ref{2b_analysis}) and
  (\ref{3b_analysis}), respectively. The numbers in parenthesis are the
  corresponding signal-to-background ratios. We show results for the
  Isajet parametrization of $b$-tagging efficiency as well as for the
  medium and tight $b$-tagging efficiencies in Ref.~\cite{CMS:2016kkf}.}
\label{tab:bcomp}
\end{table}

\section{Results}

In this section, we show that the pure sample of gluino events that we
have obtained can be used to make projections for both the gluino mass
reach as well as for the extraction of the gluino mass, along the RNS
model line introduced at the start of Sec.~\ref{sec:model}. We consider
several values of integrated luminosities at LHC14 ranging from
150~fb$^{-1}$ to the 3000~fb$^{-1}$ projected to be accumulated at the high
luminosity LHC.

\subsection{Gluino mass reach}
\label{subsec:gluino_reach}

\begin{figure}[tbp]
\includegraphics[width=\sfig,clip]{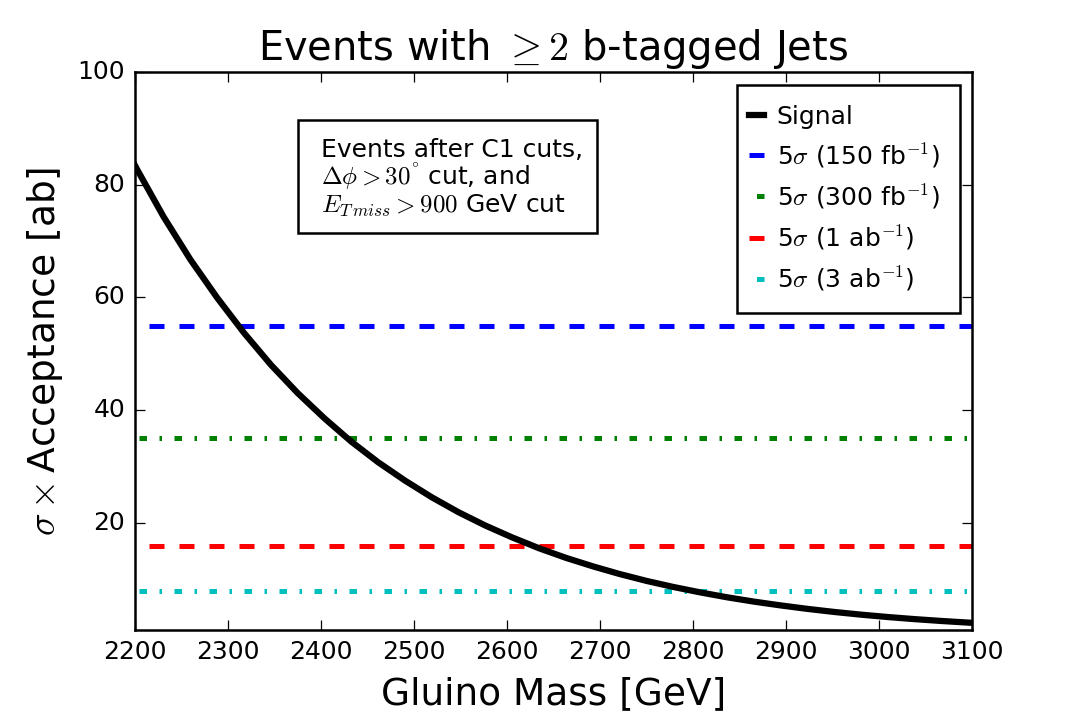}
\includegraphics[width=\sfig,clip]{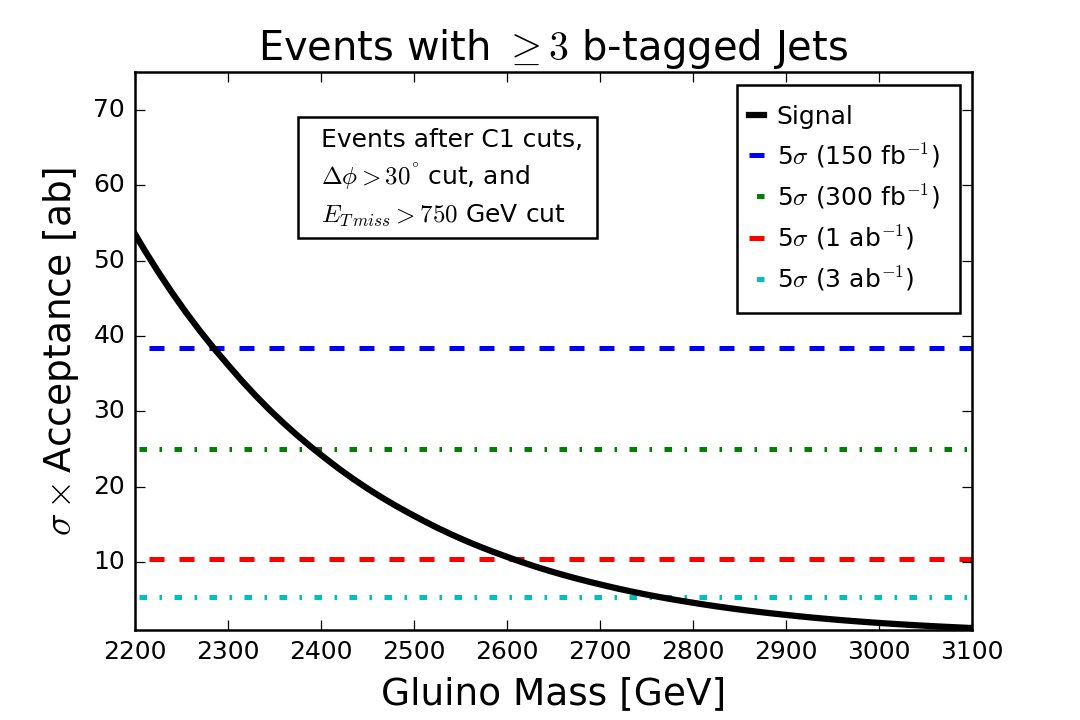}
\caption{ The gluino signal cross section for the $\ge 2$ tagged
  $b$-jet (left) and the $\ge 3$ tagged $b$-jet channels (right) after
  all the analysis cuts described in the text. The horizontal lines
  show the minimum cross section for which the Poisson fluctuation of
  the corresponding SM background levels, 5.02 ab for $2b$ events and
  1.65~ab for $3b$ events, occurs with a Gaussian probability
  corresponding to $5\sigma$ for integrated luminosities for several
  values of integrated luminosities at LHC14.
\label{fig:figreach}}
\end{figure}

We begin by showing in Fig.~\ref{fig:figreach} the gluino signal cross
section after all analysis cuts via both the $\ge 2$ tagged $b$-jets
(left frame) and the $\ge 3$ tagged $b$-jets (right frame) channels. The
total SM backgrounds in these channels are 5.02~ab and 1.65~ab,
respectively. The various horizontal lines show the minimum cross
section for which a Poisson fluctuation of the expected background
occurs with a Gaussian probability corresponding to $5\sigma$, 
for several values of integrated luminosities at
LHC14, starting with 150~fb$^{-1}$ expected (per experiment) before the
scheduled 2018 LHC shutdown, 300~fb$^{-1}$ the anticipated design
integrated luminosity of LHC14, as well as 1~ab$^{-1}$ and 3~ab$^{-1}$
that are expected to be accumulated after the high luminosity upgrade of
the LHC. We have checked that for an observable signal we always have a
minimum of five events and a sizable signal-to-background ratio.
(The lowest value for signal-to-background ratio we consider, {\it i.e.}, the
value at the maximum gluino mass for which we have $5\sigma$ discovery
with at least five events is for 3000 fb$^{-1}$ in our $2b$ analysis, for
which $S/B = 1.6$.)
We see from Fig.~\ref{fig:figreach} that, with 150~fb$^{-1}$, LHC
experiments would be probing $m_{\tg}$ values up to 2300~GeV (actually
somewhat smaller since the machine energy is still 13~TeV) via the $2b$
analysis, with only a slightly smaller reach via the $3b$ analysis. Even
for the decoupled squark scenario, we project a 3000 fb$^{-1}$ 
LHC14 $5\sigma$ gluino reach to $\sim 2400$~GeV; this will extend to
about 2800~GeV in both the $2b$ and $3b$ channels at the HL-LHC. 
These projections are significantly greater than the corresponding
reach from the mSUGRA model~\cite{lessa} because (1)~the presence of hard
$b$-jets in the signal serves as an additional handle to reduce SM
backgrounds, especially those from $W,Z$+jet production processes~\cite{btags},
and (2)~the larger $m_{\tg}-m_{\tz_1}$ mass gap expected from RNS leads to 
harder jets and harder $\eslt$ as compared to mSUGRA.
A further improvement in reach may of course be gained by combining 
ATLAS and CMS data sets.

\subsection{Gluino mass measurement} \label{subsec:gluino_mass}

We now turn to the examination of whether the clean sample of gluino
events that we have obtained allows us to extract the mass of the
gluino. For decoupled first/second generation squarks, these events can
only originate via gluino pair production. Assuming that the background
is small, or can be reliably subtracted, the event rate is completely
determined by $m_{\tg}$. A determination of this event rate after 
the analysis cuts in (\ref{2b_analysis}) or (\ref{3b_analysis}) should, in
principle, yield a measure of the gluino mass.
\begin{figure}[tbp]
\includegraphics[width=\sfig,clip]{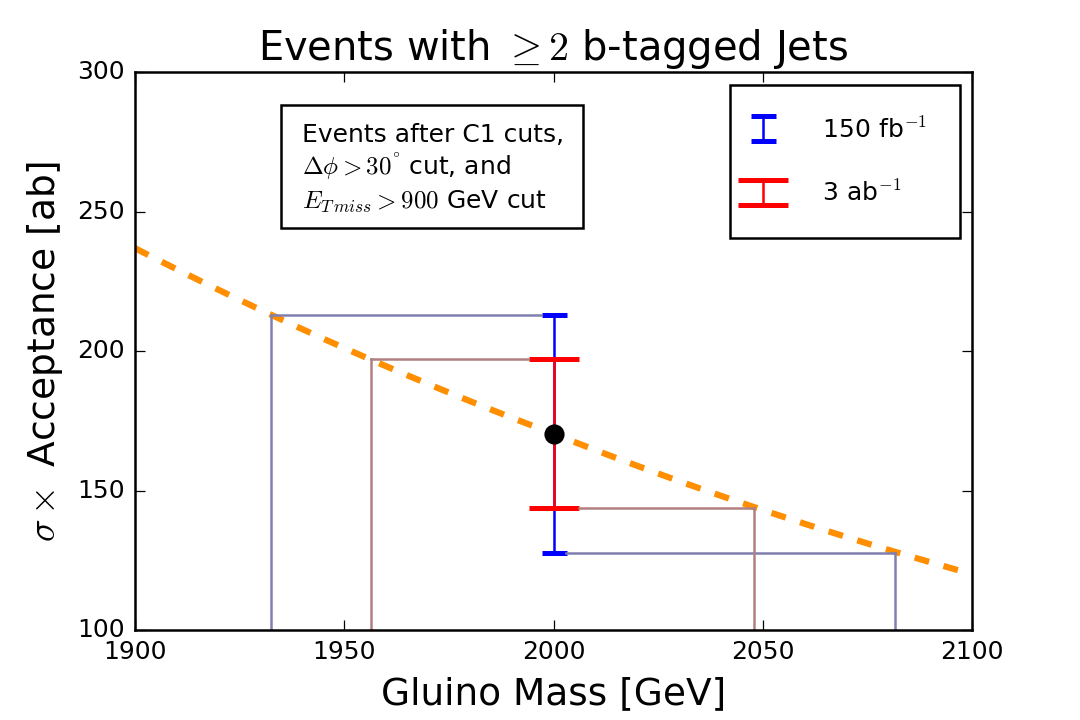}
\includegraphics[width=\sfig,clip]{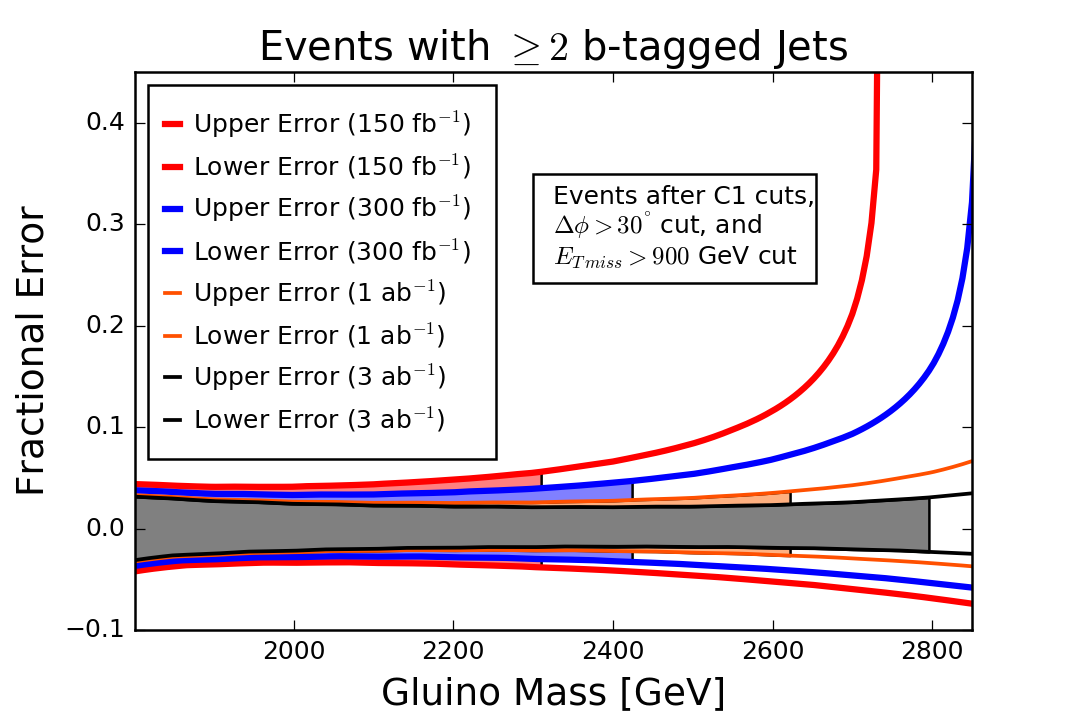}
\caption{Illustration of our method to extract the precision with which
  the gluino mass may be extracted at the LHC for the $2b$ sample (left
  frame) and the statistical precision that may be attained as a
  function of $m_{\tg}$ for integrated luminosities of 150~fb$^{-1}$,
  300~fb$^{-1}$, 1~ab$^{-1}$ and 3~ab$^{-1}$ (right frame). The left
  frame shows a blow-up of the gluino signal cross section versus
  $m_{\tg}$ for the $\ge 2$ tagged $b$-jets after all the analysis cuts described
  in the text. Also shown are the ``$1\sigma$'' error bars for a
  determination of this cross section (where the $1\sigma$ statistical error on
  the observed number of signal events and a 15\% uncertainty on the
  gluino production cross section have been combined in quadrature) for
  an integrated luminosity of 150~fb$^{-1}$ (blue) and 3~ab$^{-1}$
  (red). The other lines show how we obtain the precision with which the
  gluino mass may be extracted for our benchmark gluino point for these
  two values of integrated luminosities. The bands in the right frame
  illustrate the statistical precision on the extracted value of
  $m_{\tg}$ that may be attained at the LHC for four different values of
  integrated luminosity. We terminate the shading at the $5\sigma$
  discovery reach shown in Fig~\ref{fig:figreach}.}
\label{fig:mass2b}
\end{figure}

Our procedure for the extraction of the gluino mass (for our benchmark
point) is illustrated in the left frame of Fig.~\ref{fig:mass2b}, where
we show a blow-up of the SUSY signal cross section versus $m_{\tg}$ for
$\ge 2$~tagged $b$-jet events after all our analysis cuts. The signal
cross section can be inferred from the observed number of events in the
sample and subtracting the expected background. The error bar shown in
the figure is obtained by combining in quadrature the $1\sigma$
statistical error on the cross section based on the expected total
number of (signal plus background) events expected in the sample, with a
15\% theoretical error on the gluino production cross section. This
error bar is used to project ``the $1\sigma$'' uncertainty in the
measurement of $m_{\tg}$. From the figure, $m_{\tg}=2000^{+80}_{-70}$~GeV
with 150~fb$^{-1}$, and $m_{\tg} = (2000^{+50}_{-45})$~GeV with
3~ab$^{-1}$.  The right frame of Fig.~\ref{fig:mass2b} shows the
precision with which the gluino mass may be extracted via the clean
events in the $\ge 2$ tagged $b$-jets channel versus the gluino mass for
four different values of integrated luminosity ranging from
150~fb$^{-1}$ to 3~ab$^{-1}$. The shading on the various bands extends
out to the $5\sigma$ reach projection in Fig.~\ref{fig:figreach}. We see
that gluino mass extraction with a sub-ten percent precision is possible
with even 150~fb$^{-1}$ of integrated luminosity if gluinos are lighter
than 2.5~TeV and cascade decay via stops into light higgsinos as in the
RNS framework. It should be noted though that the $5\sigma$ reach of the
LHC extends to just $\sim 2.3$~TeV so that the determination, $m_{\tg}=
2.5$~TeV would be a mass measurement for a discovery with a significance
smaller than the customary $5\sigma$. At the high luminosity LHC, the
gluino mass may be extracted with a statistical precision better than
2.5-4\% (depending on their mass) all the way up to $m_{\tg}\sim2.8$~TeV,
{\it i.e}, if gluinos are within the $5\sigma$ discovery range of the HL-LHC! 
Gluino mass determination would also be possible for the range of
gluino masses for which the discovery significance was smaller than
$5\sigma$.

\begin{figure}[tbp]
\includegraphics[width=\sfig,clip]{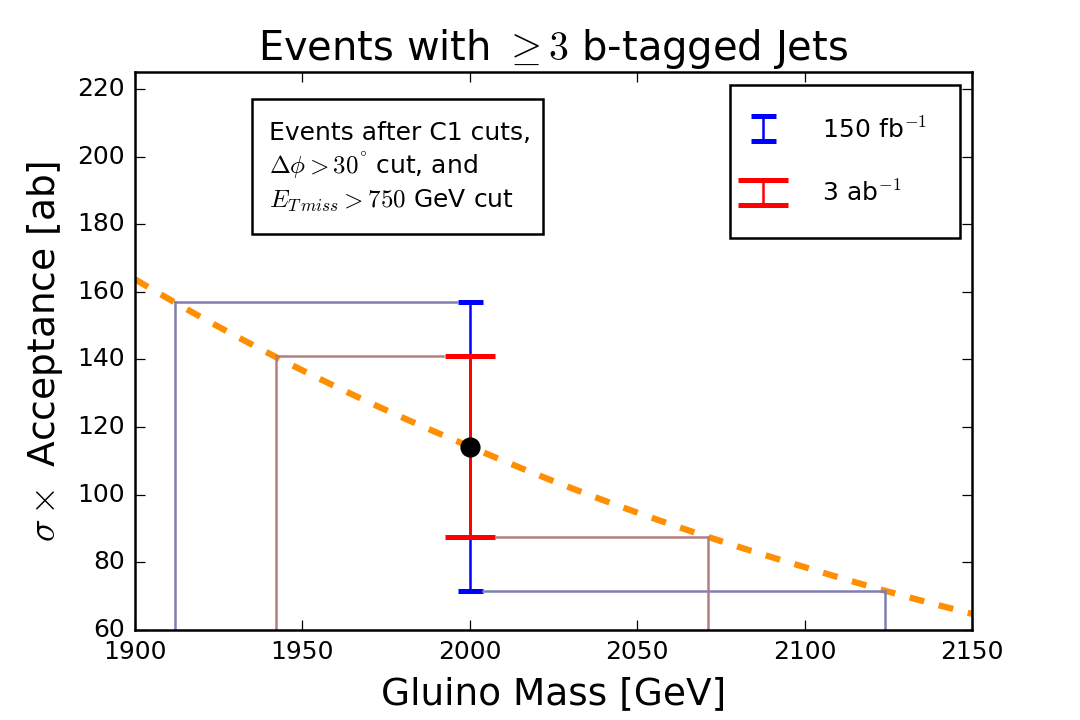}
\includegraphics[width=\sfig,clip]{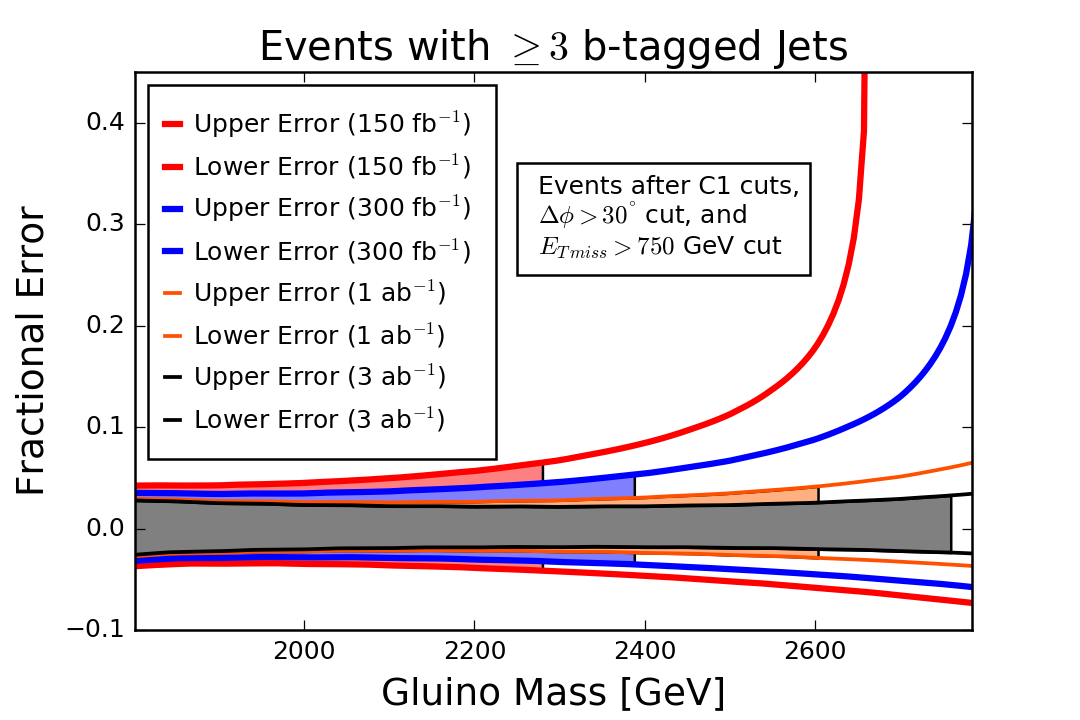}
\caption{The same as Fig.~\ref{fig:mass2b}, but for the clean SUSY
  sample with $\ge 3$ tagged $b$-jets.}
\label{fig:mass3b}
\end{figure}
Prospects for gluino mass measurement via the $\ge 3$~tagged $b$-jet
sample are shown in Fig.~\ref{fig:mass3b}. We see that the statistical
precision on the mass measurement that may be attained is somewhat worse
than that via the $\ge 2b$ channel shown in Fig.~\ref{fig:mass2b},
though not qualitatively different  except at the high mass end. The difference
is, of course, due to the  lower event rate in this channel.

Before proceeding further, we point out that in order to extract the
gluino mass, we have assumed that our estimate of the background is
indeed reliable. Since the expected background has to be subtracted
from the observed event rate to obtain the signal cross section, and
via this the value of $m_{\tg}$, any error in the estimation of the expected
background will result in a {\it systematic shift} in the extracted
gluino mass. For instance, an over-estimation of the background
expectation compared to its true value, will result in too small a
signal and a corresponding overestimate of the mass of the gluino.  We
expect that by the time a precise mass measurement becomes feasible,
it will be possible to extract the SM background to a good precision
by extrapolating the backgrounds normalized in the ``background region''
(that are expected to have low signal contamination) to the ``signal
region'' using the accumulated data.
\begin{figure}[tbp]
\begin{center}
  \includegraphics[width=\sfig,clip]{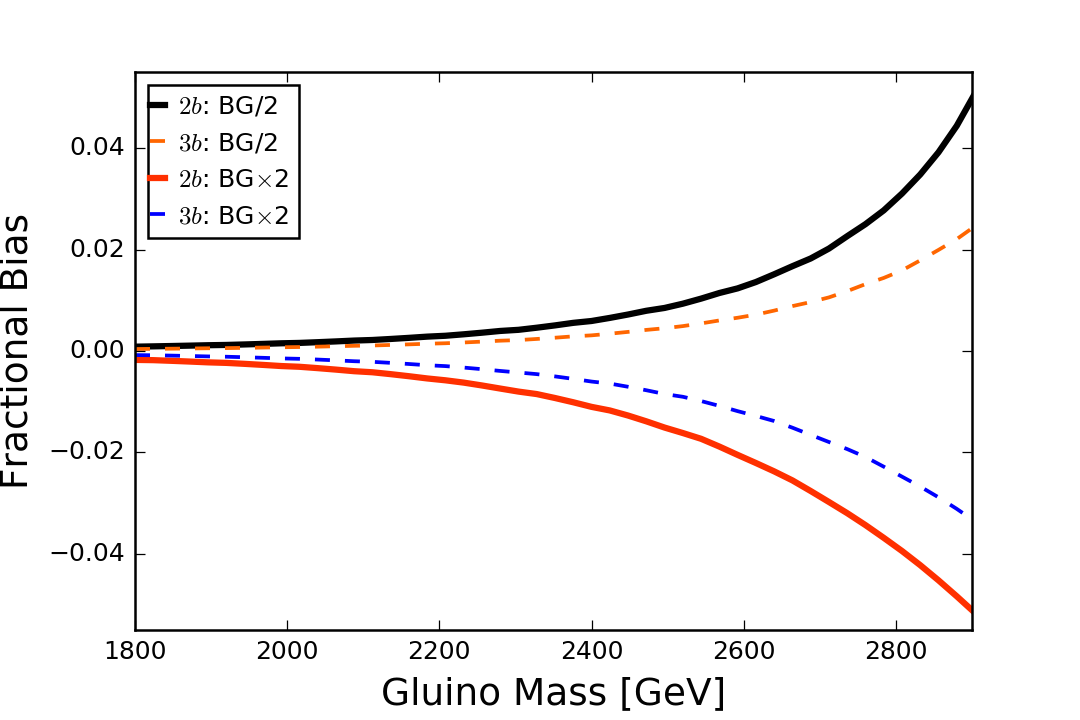}
\caption{The systematic bias, discussed in the text, in the
  measurement of the gluino mass resulting from a mis-estimate of the
  SM background by a factor of 2 in either direction. The solid lines
  are for the signal in the $2b$ channel while the dashed lines are
  for the signal in the $3b$ channel. }
\label{fig:bias}
\end{center}
\end{figure}
We show in Fig.~\ref{fig:bias} the systematic bias on the gluino mass
that could result because the background estimate differs from the
true value by a factor of 2. We see that this (asymmetric) systematic
bias is below 2\% for $m_{\tg} \alt 2.6$~TeV, but becomes as large as
4\% for the largest masses for which there is a 5$\sigma$ signal at
the high luminosity LHC in the two tagged $b$-jets sample. This bias
is smaller for the three tagged $b$-jets sample because the
corresponding background is smaller.

Our conclusions for the precision with which LHC measurements might
extract the gluino mass are very striking, and we should temper
these with some cautionary remarks. The most important thing is that any
extraction of the mass from the absolute event rate assumes an excellent
understanding of the detector in today's environment as well as in the
high luminosity environment of future experiments. While we are well
aware that our theorists' simulation does not include many important
effects, {\it e.g.}, particle detection efficiencies, jet energy scales,
full understanding of $b$-tagging efficiencies particularly for very
high $E_T$ $b$-jets, to name a few, we are optimistic that these will
all be very well understood (given that there will be a lot of data) by
the time gluino mass measurements become feasible. The fact that our
proposal relies on an inclusive cross section with $\ge 4$ jets
(of which 2 or 3 are $b$-jets) and does not entail very high jet 
multiplicities suggests that our procedure should be relatively robust. 
An excellent understanding of the
$\eslt$ tail from SM sources, as well as of the tagging efficiency (and
associated purity) for very high $E_T$ $b$-jets are crucial elements for
this analysis.

\section{Summary}\label{sec:conc}

In this paper, we have re-examined LHC signals from the pair
production of gluinos assuming gluinos decay via $\tg\to t\tst_1$,
followed by stop decays,  $\tst_1\to b\tw_1,t\tz_{1,2}$, to higgsinos,
where the visible decay products of the higgsinos are very soft.  This
is the dominant gluino decay chain expected within the
radiatively-driven natural SUSY framework that we have suggested for
phenomenological analysis of simple natural SUSY GUT models. 
For our analysis, we
have used the RNS model line detailed in Sec.~\ref{sec:model} with
higgsino masses $\sim 150$~GeV.  The
gluino signal then consists of events with $\ge 4$ hard jets, two or
three of which are tagged as $b$-jets together with very large $\eslt$.
We expect that our results are only weakly sensitive to our choice of 
higgsino mass as long as the electroweak fine-tuning parameter
$\Delta_{\rm EW} \alt 30$.

The new features that we have focussed on in this analysis are the very
large data sets (300-3000~fb$^{-1}$) that are expected to be available
at the LHC and its high luminosity upgrade and the capability for
tagging very hard $b$-jets with $E_T$ up to a TeV and beyond.  We have
identified a set of very stringent cuts, detailed in (\ref{2b_analysis})
and (\ref{3b_analysis}), that allows us to isolate the gluino signal
from SM backgrounds: our procedure yields a signal to background $> 30$
($> 60$) in the two (three) tagged $b$-jets channel for
$m_{\tg}=2$~TeV, and $>4$ ($>10$) for $m_{\tg}=2.6$~TeV. Even for
decoupled squarks, these relatively pure data samples extend the gluino
discovery reach in the RNS framework to 2.4~TeV for an integrated
luminosity of 300~fb$^{-1}$ expected by end of the current LHC run, and
to 2.8~TeV with 3000~fb$^{-1}$ anticipated after the luminosity upgrade
of the LHC. These may be compared to projections~\cite{lessa} for the
gluino reach of 1.8~TeV (2.3~TeV) for 300~fb$^{-1}$ (3000~fb$^{-1}$)
within the mSUGRA/CMSSM framework. We attribute the difference to, (1)~the
presence of $b$-jets in the signal which serve to essentially eliminate
SM backgrounds from $V$+jet production, and also reduce those from other
sources and (2)~the comparatively harder jet $E_T$ and $\eslt$ spectrum
associated with RNS models.

The separation of a relatively clean gluino sample also allows a
determination of the gluino mass based on the {\em signal event rate}
rather than kinematic properties of the event. Although the
determination of the mass from the event rate hinges upon being able to
predict the absolute normalization of the expected signal after cuts,
and so requires an excellent understanding of the detector, we are
optimistic that LHC experimenters will be able to use the
available data to be able to reliably determine acceptances and
efficiencies in the signal region by the time these measurements become
feasible. We project that with 300~fb$^{-1}$ of integrated luminosity,
experiments at LHC14 should be able to measure $m_{\tg}$ with a $1\sigma$
statistical error of $< 4.5$\% for $m_{\tg}=2.4$~TeV, {\it i.e.}, all the
way up to its $5\sigma$ discovery limit. At the high luminosity LHC, the
projected precision for a gluino mass measurement ranges between about
2.5\% for $m_{\tg}=2$~TeV to about 4\% near its $5\sigma$ discovery limit
of 2.8~TeV in the $\ge 2$ tagged $b$-jet channels. Comparable precision
is obtained also via the $\ge 3$ tagged $b$ channel. In this connection,
we should also keep in mind that a factor of two uncertainty in the
projected background will result in a small (but not negligible)
systematic uncertainty ranging between $ <1$\% for $m_{\tg}< 2500$~GeV
to about 3\% for $m_{\tg}= 2800$ GeV in the extraction of $m_{\tg}$ via
the $\ge 2$ tagged $b$ channel and smaller than this for the $\ge 3$
tagged $b$ channel.

An observation of a SUSY signal in both $2b$ and $3b$ channels, and the
extraction of a common value of $m_{\tg}$ would certainly be strong
evidence for a discovery of a new particle. If these signals are
accompanied by other signals such as the same-sign diboson signal
and/or monojet events with soft opposite sign dileptons, the case for
the discovery of radiatively-driven natural SUSY would be very strong. 
In this case, depending on the gluino mass, 
there may also be signals in trilepton and even four
lepton plus jets plus $\eslt$ channels~\cite{lhc}.

\begin{figure}[tbp]
\begin{center}
  \includegraphics[width=15cm,clip]{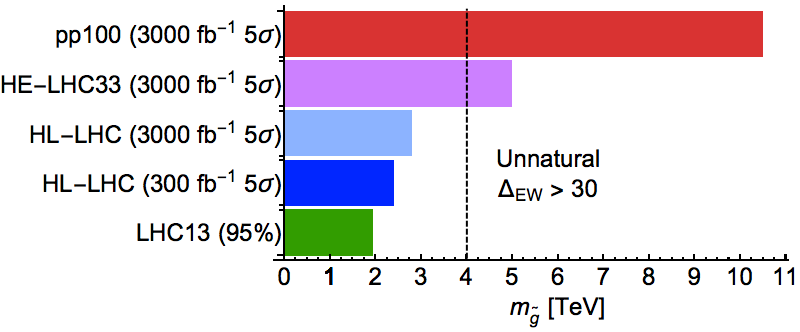}
\caption{The approximate reach for various present and future hadron
collider options for gluino pair production. The region to the right of
the dashed line yields large electroweak fine-tuning and is considered
unnatural.  }
\label{fig:bar}
\end{center}
\end{figure}

In Fig. \ref{fig:bar} we compare the approximate reach for various
present and future hadron collider options for gluino pair production.
The region to the right of the dashed line yields large electroweak
fine-tuning and is considered unnatural.  The green bar shows the
present LHC 95\% CL limit on $m_{\tg}$ as derived in several simplified
models which should be applicable to the present RNS case.  The dark and
light blue bars show our projected LHC14 300 and 3000 fb$^{-1}$
$5\sigma$ reaches for RNS. These cover only a portion of natural SUSY
parameter space. The lavendar bar shows the reach of HE-LHC with
$\sqrt{s}=33$ TeV as abstracted from Ref. \cite{Gershtein:2013iqa} where
it is assumed that the gluino directly decays to a light LSP via $\tg\to
q\bar{q}\tz_1$ (presumably with no enhancement of decays to third
generation quarks). The $5\sigma$ HE-LHC for 3000 fb$^{-1}$ extends to
$m_{\tg}\sim 5$ TeV and thus covers {\it all} of natural SUSY parameter
space.  The red bar shows the corresponding gluino reach of a 100 TeV
$pp$ collider at $5\sigma$ and 3000 fb$^{-1}$, as taken also from
Ref. \cite{Gershtein:2013iqa}.  Here, the reach extends just beyond
$m_{\tg}\sim 10$ TeV. It probes only more deeply into unnatural SUSY
parameter space beyond the complete coverage of the gluino offered by
HE-LHC, but does offer the possibility of a squark discovery.

In summary, in models such as RNS where gluinos dominantly decay via
$\tg\to t\tst_1$, and the stops decay to light higgsinos via $\tst_1 \to
\tw_1b$, $\tz_{1,2}t$, signals from gluino pair production should be
observable at the $5\sigma$ level out to $m_{\tg} < 2.4 \ (2.8)$~TeV for
an integrated luminosity of 300~fb$^{-1}$ (3000~fb$^{-1}$) in the $\ge
4$-jet sample with very hard $\eslt$ and two or three tagged $b$-jets. 
The clean sample of gluino events that we obtain should also
allow a measurement of $m_{\tg}$ with a statistical precision ranging
from 2.5-4\% depending on the gluino mass and the integrated luminosity,
along with a smaller but non-negligible systematic uncertainty of 1-3\%
mentioned in the previous paragraph.
The precision of gluino mass extraction should be even greater 
using the combined ATLAS/CMS data set.

\section*{Acknowledgments}

We thank Heather Gray, Jacobo Konigsberg, Bill Murray, Markus Klute
and Giacomo Polesello for helpful discussions about $b$-jet 
tagging at the LHC and we thank P. Skubic for reading the manuscript.
This work was supported
in part by the US Department of Energy, Office of High Energy Physics,
and was aided by the use of SLAC Computing Resources and the
LHCO\_reader Python package~\cite{Fowlie:2015dga}.  JG would like to
express a special thanks to the Mainz Institute for Theoretical
Physics (MITP) for its hospitality and support.
VB thanks the KITP at University of California-Santa Barbara for hospitality.  
This research was supported in part by the National Science Foundation 
under Grant No. NSF PHY11-25915.
%

\end{document}